\documentclass[a4paper]{article}

 \pdfoutput=1
\usepackage{color}
\usepackage{graphicx}
\usepackage{amsfonts}
\usepackage{amssymb}
\usepackage{verbatim}
\usepackage[verbose,a4paper,tmargin=1.0in,bmargin=1.0in,lmargin=1.0in,rmargin=1.0in,nohead]{geometry}
\usepackage[super,sort&compress]{natbib}

\bibpunct{}{}{,}{s}{}{;}

\newcommand{\NPsection}[1]{\subsection*{\textsf{\normalsize\color{black}#1}}}

\usepackage{setspace}
\usepackage{hyperref}
\hypersetup{pdfauthor={Matteo Mariantoni},pdftitle={Photon shell game in three-resonator circuit quantum
  electrodynamics}}

\begin{document}

\renewcommand{\topfraction}{1.0}
\renewcommand{\bottomfraction}{1.0}
\renewcommand{\textfraction}{0.0}

\noindent\parbox{\textwidth}{\flushleft\textsf{\color{black}\Huge
Photon shell game in three-resonator circuit quantum
electrodynamics}}
    \vspace{3mm}

\noindent\parbox{\textwidth}{\flushleft
\textsf{\color{black}\Large
        Matteo~Mariantoni$^{\mathsf{1,3,*}}$,
        H.~Wang$^{1}$,
        Radoslaw~C.~Bialczak$^{1}$,
        M.~Lenander$^{1}$,
        Erik~Lucero$^{1}$,
        M.~Neeley$^1$,
        A.~D.~O'Connell$^{1}$,
        D.~Sank$^{1}$,
        M.~Weides$^{1}$,
        J.~Wenner$^{1}$,
        T.~Yamamoto$^{1,2}$,
        Y.~Yin$^{1}$,
        J.~Zhao$^1$,
        John~M.~Martinis$^{1,\dag}$
        and A.~N.~Cleland$^{1,\ddag}$
       }
    \vspace{2mm}

\textsf{\textbf{\small $^{1}$Department of Physics, University of California, Santa Barbara, California 93106, USA\\
$^{2}$Green Innovation Research Laboratories, NEC Corporation, Tsukuba, Ibaraki 305-8501, Japan\\
$^{3}$California NanoSystems Institute, University of California, Santa Barbara, California 93106, USA\\
$^{*}$e-mail: matmar@physics.ucsb.edu\\
$^{\dag}$e-mail: martinis@physics.ucsb.edu\\
$^{\ddag}$e-mail: anc@physics.ucsb.edu
    \vspace{5mm}
    }}}

\noindent\textsf{\small last updated: \today}\\
{\color{black}\rule[2mm]{\textwidth}{0.1mm}}



    {\noindent\textbf{
The generation and control of quantum states of light
constitute fundamental tasks in cavity quantum electrodynamics
(QED)~\cite{mabuchi:2002:a,haroche:2006:a,walther:2006:a,
rauschenbeutel:2001:a,gleyzes:2007:a,deleglise:2008:a,
hijlkema:2007:a,wilk:2007:a,dayan:2008:a,papp:2009:a}. The
superconducting realization of cavity QED, circuit
QED~\cite{wallraff:2004:a,chiorescu:2004:a,johansson:2006:a,schoelkopf:2008:a},
enables on-chip microwave photonics, where superconducting
qubits~\cite{makhlin:2001:a,wendin:2006:a,you:2005:a,clarke:2008:a}
control and measure individual photon
states~\cite{houck:2007:a,sillanpaa:2007:a,hofheinz:2008:a,deppe:2008:a,
hofheinz:2009:a,wang:2009:a,leek:2010:a,johnson:2010:a}. A
long-standing issue in cavity QED is the coherent transfer of
photons between two or more resonators. Here, we use circuit
QED to implement a three-resonator architecture on a single
chip, where the resonators are interconnected by two
superconducting phase qubits. We use this circuit to shuffle
one- and two-photon Fock states between the three resonators,
and demonstrate qubit-mediated vacuum Rabi swaps between two
resonators. This illustrates the potential for using
multi-resonator circuits as photon quantum registries and for
creating multipartite entanglement between delocalized bosonic
modes~\cite{mariantoni:2008:a}.
    }

The combination of high-finesse electromagnetic cavities with
atoms or qubits enables fundamental studies of the interaction
between light and matter. The cavity provides a protected
environment for storing and tailoring individual photonic
excitations~\cite{mabuchi:2002:a,haroche:2006:a,walther:2006:a,schoelkopf:2008:a}.
Both stationary~\cite{deleglise:2008:a,hofheinz:2009:a} and
propagating~\cite{hijlkema:2007:a,dayan:2008:a} nonclassical
fields can be synthesized using such systems, enabling quantum
memory and quantum messaging~\cite{wilk:2007:a}. A critical
challenge however is the extension from single to more
versatile multi-cavity
architectures~\cite{mariantoni:2008:a,helmer:2009:a}, allowing
manipulation of spatially separated bosonic modes. While the
entanglement of different modes of a single cavity has been
shown in atomic
systems~\cite{rauschenbeutel:2001:a,papp:2009:a}, and a coupled
low- and high-quality factor resonator studied in circuit
QED~\cite{leek:2010:a,johnson:2010:a}, coherent dynamics
between two or more high-quality factor cavities has yet to be
demonstrated. Here we describe a triple-resonator system, where
three high-quality factor microwave resonators are coupled to
two superconducting phase qubits
(cf.~Fig.~\ref{Figure1MatteoMariantoni201007}). The qubits
serve as quantum transducers that create and transfer photonic
states between the resonators~\cite{sun:2005:a}. The quantum
transduction is performed by means of purely \textit{resonant}
qubit-resonator interactions, rather than dispersive
coupling~\cite{mariantoni:2008:a}, enabling rapid transfers
between resonators with significantly different frequencies. As
an important example, we demonstrate single-photon Rabi swaps
between two resonators de-tuned by $\simeq {} 12000$ resonator
linewidths.

\begin{figure}[t!]
    \centering
    \includegraphics[width=0.99\columnwidth]{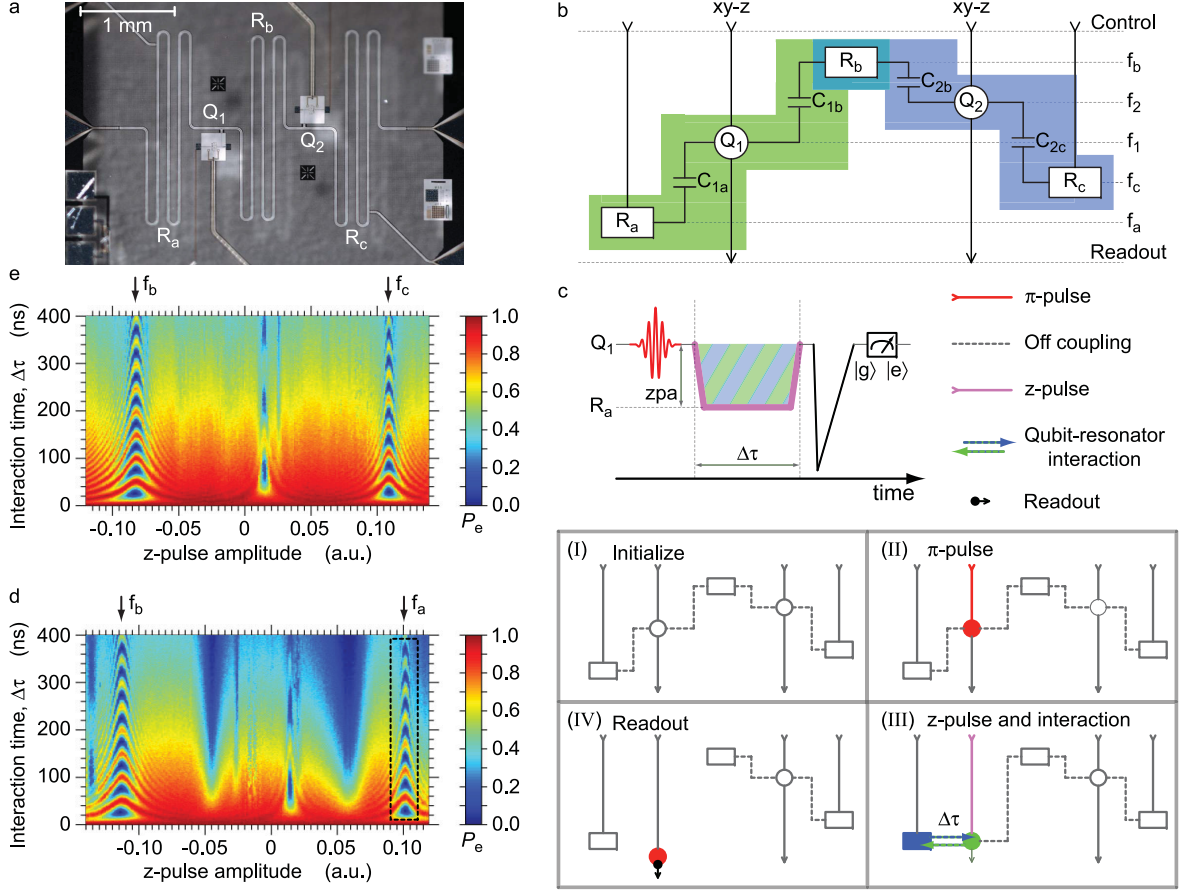}
    \caption{\footnotesize
\textbf{Experimental architecture and two-dimensional swap
spectroscopy.} \textbf{a,} Optical micrograph of sample,
showing three coplanar waveguide resonators
(R$^{}_{\textrm{a}}$, R$^{}_{\textrm{b}}$ and
R$^{}_{\textrm{c}}$, with meander design) capacitively coupled
to two superconducting phase qubits (Q$^{}_1$ and Q$^{}_2$, in
boxes). \textbf{b,} Block diagram showing main elements, which
comprise two circuit unit cells
R$^{}_{\textrm{a}}$-Q$^{}_1$-R$^{}_{\textrm{b}}$ (light green
area) and R$^{}_{\textrm{b}}$-Q$^{}_2$-R$^{}_{\textrm{c}}$
(dark blue area). The horizontal placement represents the
spatial layout of the sample, whereas the vertical distribution
corresponds to the frequencies of the elements. The
qubit-resonator coupling capacitors are designed to be $C_{1
\textrm{a}} {} = {} C_{1 \textrm{b}} {} = {} C_{2 \textrm{b}}
{} = {} C^{}_{2 \textrm{c}} {} = {} 1.9$\,fF. $f^{}_{\rm a} {}
\simeq {} 6.29$\,GHz, $f^{}_{\rm b} {} \simeq {} 6.82$\,GHz and
$f^{}_{\rm c} {} \simeq {} 6.34$\,GHz are the measured
resonator frequencies, and $f^{}_1$ and $f^{}_2$ the tuneable
qubit transition frequencies. Control and readout wiring is
also shown. \textbf{c,} Upper sub-panel, pulse sequence for
swap spectroscopy, with data shown in \textbf{d} and
\textbf{e}. $Q^{}_1$ line shows Gaussian microwave $\pi$-pulse
(red) and qubit tuning pulse (z-pulse; hashed magenta) with
variable $z$-pulse amplitude {zpa} and duration $\Delta \tau$,
followed by a triangular measurement pulse (black). Lower
sub-panel, diagrammatic representation of the pulse sequence.
(I) The entire system is initialized in the ground state. (II)
Q$^{}_1$ is excited by a $\pi$-pulse (red), then (III) brought
into resonance with R$^{}_{\textrm{a}}$ via a $z$-pulse
(magenta) and allowed to interact with the resonator for a time
$\Delta \tau$. (IV) A measurement pulse projects the qubit onto
its ground state $| \textrm{g} \rangle$ or excited state $|
\textrm{e} \rangle$. \textbf{d,} Two-dimensional swap
spectroscopy for Q$^{}_1$. The probability $P^{}_{\textrm{e}}$
to find the qubit in $| \textrm{e} \rangle$ is plotted versus
$z$-pulse amplitude and resonator measurement time $\Delta
\tau$. The typical chevron pattern generated by a
qubit-resonator swap (arrows) is evident for both the
Q$^{}_1$-R$^{}_{\textrm{a}}$ (dashed black box) and
Q$^{}_1$-R$^{}_{\textrm{b}}$ interactions. Near the center of
the plot a qubit interaction with a spurious two-level system
is seen, surrounded by two regions with short qubit relaxation
time. \textbf{e,} Same as \textbf{d} for Q$^{}_2$. From these
measurements, we find the coupling strengths $g^{}_{1 {\rm a}}
{} \simeq {} 17.58 {} \mp {} 0.01$\,MHz, $g^{}_{1 {\rm b}} {}
\simeq {} 20.65 {} \mp {}0.02$\,MHz, $g^{}_{2 {\rm b}} {}
\simeq {} 20.43 {} \mp {} 0.01$\,MHz and $g^{}_{2 {\rm c}} {}
\simeq {} 17.96 {} \mp {} 0.01$\,MHz.
    }
    \label{Figure1MatteoMariantoni201007}
\end{figure}

Figure~\ref{Figure1MatteoMariantoni201007}a,b shows the main
experimental elements, which comprise three coplanar waveguide
resonators, R$^{}_{\textrm{a}}$, R$^{}_{\textrm{b}}$ and
R$^{}_{\textrm{c}}$, two phase qubits, Q$^{}_1$ and Q$^{}_2$,
and two superconducting quantum interference devices (SQUIDs)
used for qubit state readout. Each qubit is coupled to a
control line which is used to adjust the qubit operating
frequency $f^{}_{1 , 2}$ and couple microwave pulses for
controlling and measuring the qubit state. During operation,
the device is attached to the mixing chamber of a dilution
refrigerator at $\approx {} 25$\,mK.

\begin{figure}[t!]
    \centering
    \includegraphics[width=0.98\columnwidth]{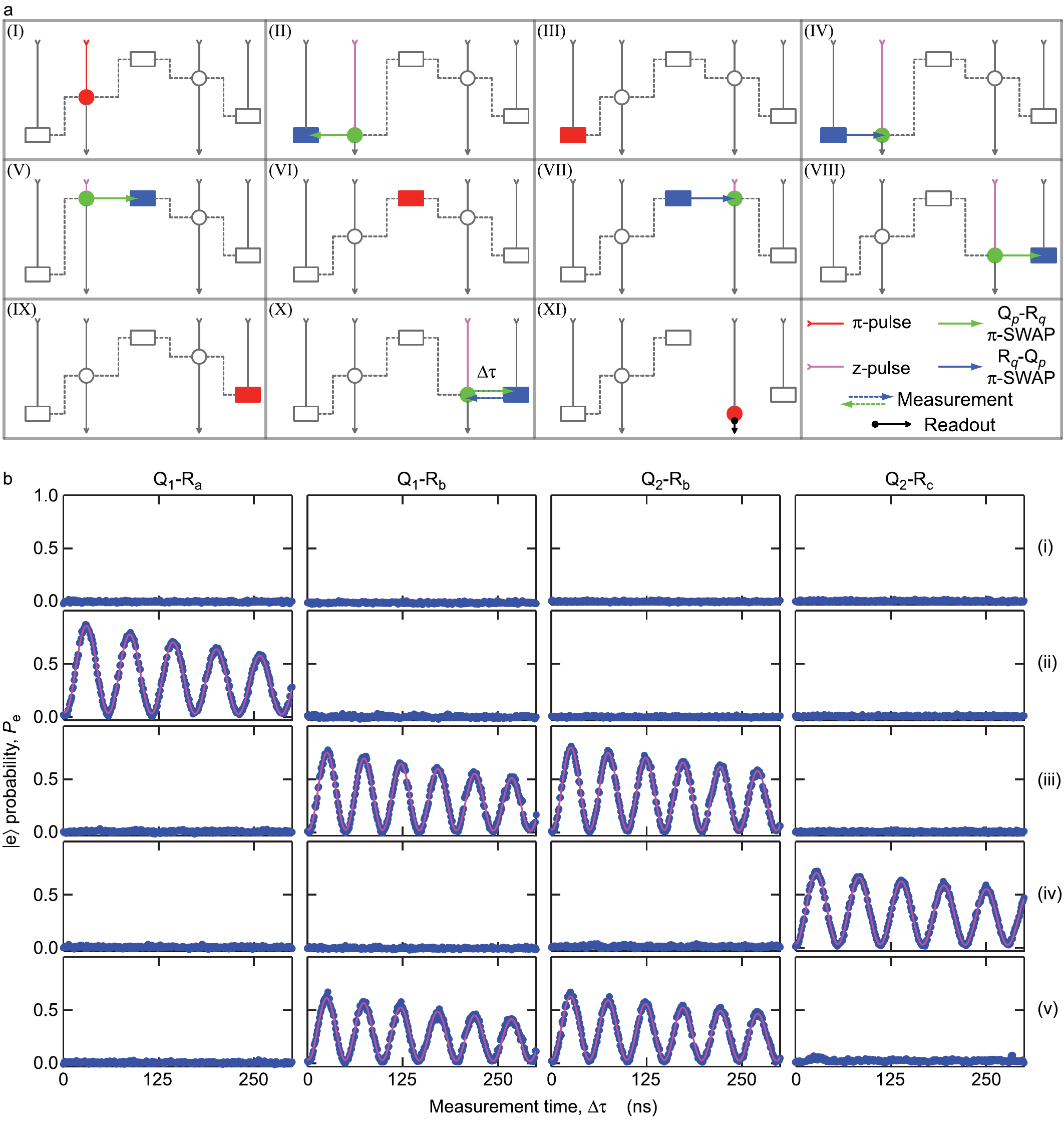}
    \caption{\footnotesize
\textbf{Photon shell game.} \textbf{a,} Block diagram of the
sequence used to coherently transfer a single photon Fock state
$| 1 \rangle$ from R$^{}_{\textrm{a}}$ to R$^{}_{\textrm{c}}$
via R$^{}_{\textrm{b}}$. After initializing the system in the
ground state, (I), Q$^{}_1$ is excited by a $\pi$-pulse and,
(II), $z$-pulsed into resonance with R$^{}_{\textrm{a}}$ for a
full Rabi swap (Rabi $\pi$-swap) at the end of which, (III),
R$^{}_{\textrm{a}}$ is populated by the one-photon Fock state
$| 1 \rangle$ and Q$^{}_1$ is in its ground state, at the idle
point. In (IV), Q$^{}_1$ is $z$-pulsed into resonance with
R$^{}_{\textrm{a}}$ for a Rabi $\pi$-swap and then, (V),
$z$-pulsed into resonance with R$^{}_{\textrm{b}}$ for another
Rabi $\pi$-swap at the end of which, (VI), R$^{}_{\textrm{b}}$
is populated by the one-photon Fock state and both Q$^{}_1$ and
Q$^{}_2$ are in the ground state at the idle point. (VII),
Q$^{}_2$ is $z$-pulsed into resonance with R$^{}_{\textrm{b}}$
for a Rabi $\pi$-swap and, (VIII), $z$-pulsed into resonance
with R$^{}_{\textrm{c}}$ for another Rabi $\pi$-swap at the end
of which, (IX), R$^{}_{\textrm{c}}$ is in the one-photon Fock
state and Q$^{}_2$ in the ground state at the idle point.
Measurement and qubit state readout are performed in (X) and
(XI), respectively, where the presence of the Fock state in
R$^{}_{\textrm{c}}$ is detected by its interaction with
Q$^{}_2$. \textbf{b,} Measurement outcomes for different photon
shell games. Each plot shows the probability
$P^{}_{\textrm{e}}$ to measure a qubit in the excited state $|
\textrm{e} \rangle$ as a function of the qubit-resonator
measurement time $\Delta \tau$. Blue circles are data, magenta
lines a least-squares fit. Data for Q$^{}_1$ are in the first
two columns, for Q$^{}_2$ in the second two; each row
corresponds to a different game. Row (i), all resonators are in
the vacuum state (with fidelity $\mathcal{F} {} > {} 0.99$; we
define $\mathcal{F}$ as the amplitude of the fit; see
Supplementary Information), i.e. no stored photons. Row (ii),
resonator R$^{}_{\textrm{a}}$ contains one photon ($\mathcal{F}
{} = {} 0.86 {} \mp {} 0.01$), with the other two resonators in
the vacuum state. In row (iii) the photon has been placed in
R$^{}_{\textrm{b}}$ ($\mathcal{F} {} = {} 0.80 {} \mp {}
0.01$), and in (iv) the photon is in resonator
R$^{}_{\textrm{c}}$ ($\mathcal{F} {} = {} 0.69 {} \mp {}
0.01$). In (v), we have taken the photon from
R$^{}_{\textrm{c}}$ and placed it back in R$^{}_{\textrm{b}}$
($\mathcal{F} {} = {} 0.61 {} \mp {} 0.01$), demonstrating the
high degree of control and coherence in the system.
    }
    \label{Figure2MatteoMariantoni201007}
\end{figure}

The circuit layout (Fig.~\ref{Figure1MatteoMariantoni201007}b)
can be decomposed into two unit cells,
R$^{}_{\textrm{a}}$-Q$^{}_1$-R$^{}_{\textrm{b}}$ (green area)
and R$^{}_{\textrm{b}}$-Q$^{}_2$-R$^{}_{\textrm{c}}$ (blue
area). The shared resonator R$^{}_{\textrm{b}}$ connects the
two cells and protects the two qubits from unwanted crosstalk.
The resonator frequencies $f^{}_{\textrm{a}}$,
$f^{}_{\textrm{b}}$ and $f^{}_{\textrm{c}}$ are measured with
qubit spectroscopy (not shown). The vacuum Rabi couplings
between each qubit and its corresponding resonators, $g^{}_{1
\textrm{a}}$ and $g^{}_{1 \textrm{b}}$ for Q$^{}_1$ and
$g^{}_{2 \textrm{b}}$ and $g^{}_{2 \textrm{c}}$ for Q$^{}_2$,
are determined by their respective coupling capacitors
(cf.~Fig.~\ref{Figure1MatteoMariantoni201007}b). The coupling
strengths are measured using two-dimensional swap spectroscopy
(cf.~Fig.\ref{Figure1MatteoMariantoni201007}d,e). We note that
the swap spectroscopy provides an excellent tool for revealing
the presence of spurious two-level systems (TLSs) as well as
frequencies with short qubit relaxation times. In all the
experiments, the qubits are initialized in the ground state $|
\textrm{g} \rangle$ and are typically tuned to the idle point,
where the qubit Q$^{}_1$ (Q$^{}_2$) $| \textrm{g} \rangle {}
\leftrightarrow {} | \textrm{e} \rangle$ transition frequency
$f^{}_1$ ($f^{}_2$) is set in-between, and well away from, the
resonator transition frequencies $f^{}_{\textrm{a}}$ and
$f^{}_{\textrm{b}}$ ($f^{}_{\textrm{b}}$ and
$f^{}_{\textrm{c}}$).

\begin{figure}[t!]
    \centering
    \includegraphics[width=0.99\columnwidth]{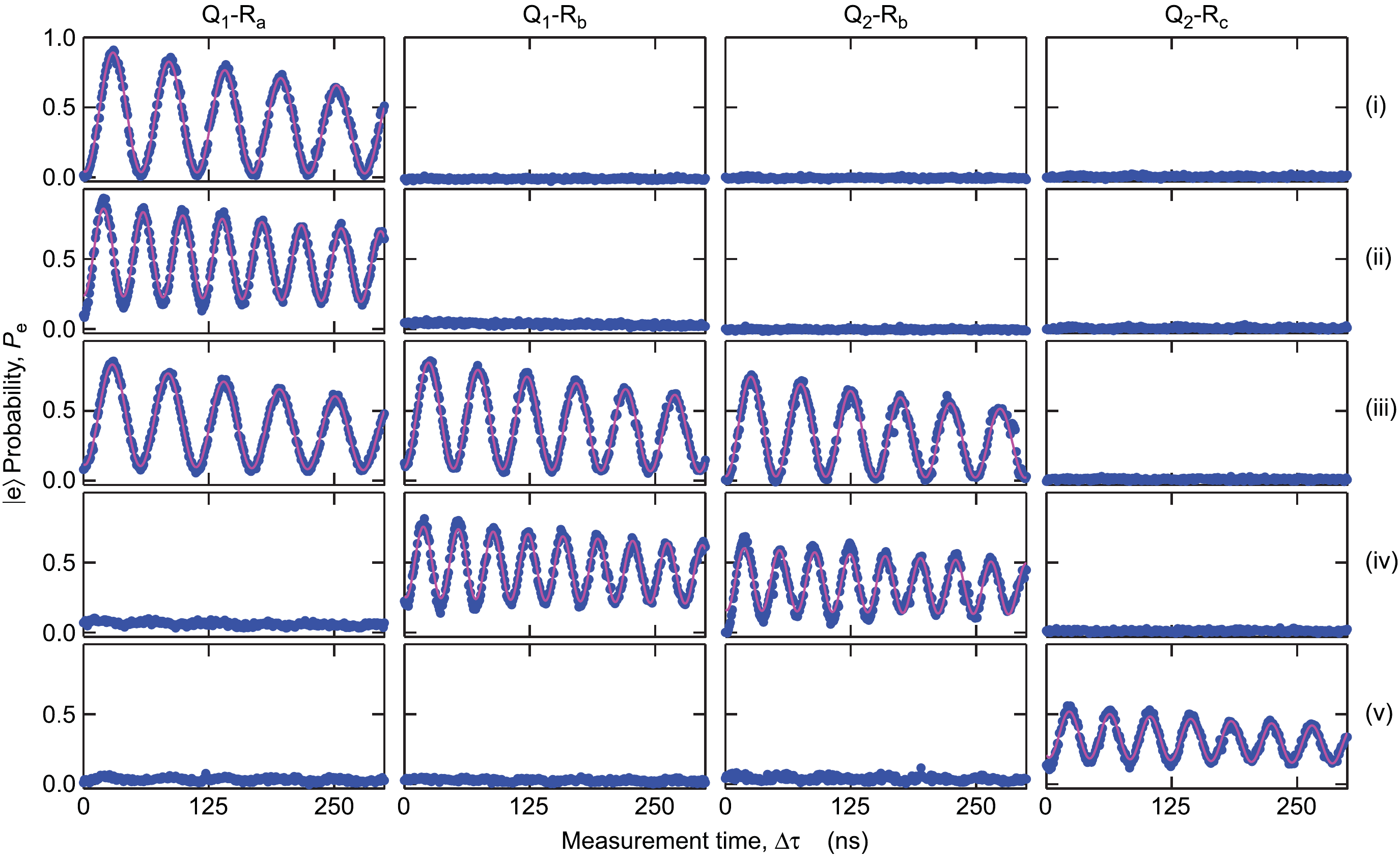}
    \caption{\footnotesize
\textbf{Quantum-mechanical realization of the `Towers of
Hanoi'.} Format as in
Fig.~\ref{Figure2MatteoMariantoni201007}b, showing the
probability $P^{}_{\textrm{e}}$ of measuring Q$^{}_1$ or
Q$^{}_2$ in the excited state as a function of interaction time
$\Delta \tau$ with a resonator. Data are shown as blue circles,
with a least-squares fit as magenta solid lines. (i), A
one-photon Fock state $| 1 \rangle$ in R$^{}_{\textrm{a}}$ with
both R$^{}_{\textrm{b}}$ and R$^{}_{\textrm{c}}$ in the vacuum
state, and (ii) a two-photon Fock state in R$^{}_{\textrm{a}}$
(fidelity $\mathcal{F} {} = {} 0.63 {} \mp {} 0.01$) with the
other two resonators in the vacuum state. A qubit-resonator
interaction oscillation is $\sqrt{2}$ faster for a two-photon
state compared to a one-photon state. In (iii), one photon has
been transferred from R$^{}_{\textrm{a}}$ to
R$^{}_{\textrm{b}}$, so one-photon oscillations are seen when
Q$^{}_1$ measures R$^{}_{\textrm{a}}$ or R$^{}_{\textrm{b}}$,
and when Q$^{}_2$ measures R$^{}_{\textrm{b}}$. In (iv), the
second photon has been transferred to R$^{}_{\textrm{b}}$
($\mathcal{F} {} = {} 0.51 {} \mp {} 0.01$), yielding the
$\sqrt{2}$ increase when either Q$^{}_1$ or Q$^{}_2$ measure
R$^{}_{\textrm{b}}$. In (v), both photons have been transferred
to R$^{}_{\textrm{c}}$ ($\mathcal{F} {} = {} 0.34 {} \mp {}
0.01$). Note that even with this complex protocol, both
R$^{}_{\textrm{a}}$ and R$^{}_{\textrm{b}}$ display negligible
oscillations ($\mathcal{F} {} \gtrsim {} 0.97$). See
Supplementary Information for further analysis.
    }
    \label{Figure3MatteoMariantoni201007}
\end{figure}

When Q$^{}_1$ (Q$^{}_2$) is at the idle point, the
qubit-resonator detuning is sufficiently large that the
qubit-resonator interactions are effectively switched off. A
particular qubit-resonator Q$^{}_p$-R$^{}_q$ ($p {} = {} 1 , 2$
and $q {} = {} \textrm{a} , \textrm{b} , \textrm{c}$)
interaction is switched on by shifting the qubit transition
frequency $f^{}_p$ to equal the resonator frequency $f^{}_q$,
setting the de-tuning to zero and enabling quantum energy
transfers. The time-dependent control of the qubit transition
frequency thus enables highly complex quantum control of the
resonators~\cite{hofheinz:2009:a}.

Figure~\ref{Figure2MatteoMariantoni201007}a shows a diagram of
the pulse sequence used to implement the single photon
equivalent of the `shell game', in which a pea is hidden under
one of three shells and the contestant must guess where the pea
is after the shells have been shuffled. The three resonators
play the role of the shells and a single photon Fock state $| 1
\rangle$ that of the pea. The system is initialized in the
ground state, with the qubits at their idle points, so that all
interactions are effectively switched off. Qubit Q$^{}_1$ is
used to pump a single photon into resonator R$^{}_{\textrm{a}}$
[Figure~\ref{Figure2MatteoMariantoni201007}a(I-III)]. The
photon state can then be transferred to either of the other two
resonators, using the qubits in a similar fashion to mediate
the single-excitation transfer. A transfer from
R$^{}_{\textrm{a}}$ to R$^{}_{\textrm{b}}$ is shown in steps
(IV-VI), and a second transfer from R$^{}_{\textrm{b}}$ to
R$^{}_{\textrm{c}}$ performed using Q$^{}_2$ in steps (VII-IX).
The final location of the photon can be determined by employing
the qubits as photon detectors, through the Rabi oscillations
that occur when a qubit is brought in resonance with a
resonator storing a photons [steps (X-XI)].

\begin{figure}[t!]
    \centering
    \includegraphics[width=0.495\columnwidth]{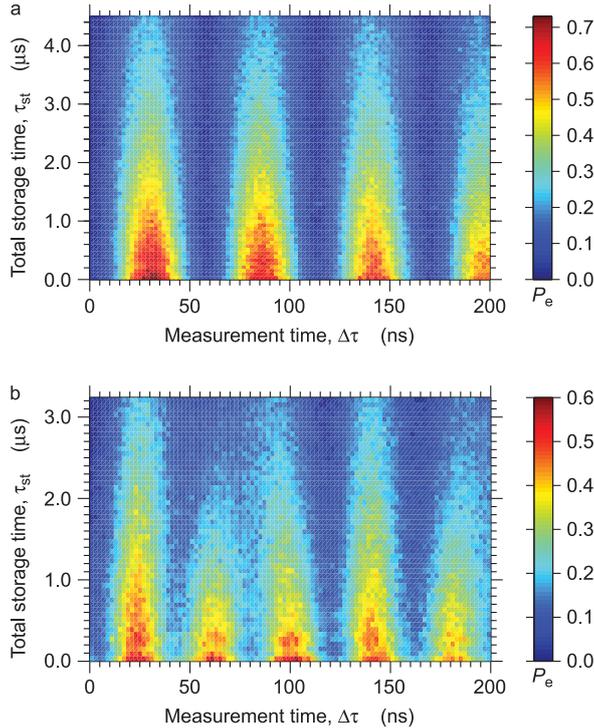}
    \caption{\footnotesize
\textbf{Combined quantum state transfer and storage for one-
and two-photon states.} \textbf{a,} Probability
$P^{}_{\textrm{e}}$ to find Q$^{}_2$ in the excited state
(colour bar scale, right side) versus measurement time $\Delta
\tau$ (horizontal axis), and total storage time
$\tau^{}_{\textrm{st}}$ (vertical axis). A one-photon Fock
state is created and stored in resonator R$^{}_{\textrm{a}}$
for a time $\tau^{}_{\textrm{st}} / 3$, transferred to
R$^{}_{\textrm{b}}$ and stored for the same time, then
transferred to R$^{}_{\textrm{c}}$, stored for the same time
and then measured. \textbf{b,} Same as in \textbf{a}, but for
the generation and storage of a two-photon Fock state.
    }
    \label{Figure4MatteoMariantoni201007}
\end{figure}

In the data shown in
Figure~\ref{Figure2MatteoMariantoni201007}b(i)-(v), which
represents different versions of the game, a photon was stored
in one of the three resonators, shuffled between the
resonators, and all three resonators then measured. In the
shell game of Fig.~\ref{Figure2MatteoMariantoni201007}b(i), no
photon was placed in any resonator, while for example in game
(iv), a single photon was transferred from R$^{}_{\textrm{a}}$
to R$^{}_{\textrm{c}}$ via R$^{}_{\textrm{b}}$, and then
detected using qubit $\textrm{Q}^{}_2$; measurements of the
other resonators R$^{}_{\textrm{a}}$ and R$^{}_{\textrm{b}}$
show no oscillations, i.e. no photonic excitation.

We also explored a variant of the shell game, transferring a
two-photon Fock state $| 2 \rangle$ from R$^{}_{\textrm{a}}$ to
R$^{}_{\textrm{b}}$ to R$^{}_{\textrm{c}}$. The two-photon Fock
state is first generated in
R$^{}_{\textrm{a}}$~\cite{hofheinz:2008:a}, as shown by the
measurements in
Fig.~\ref{Figure3MatteoMariantoni201007}(i),(ii).
Figure~\ref{Figure3MatteoMariantoni201007} also shows the
measurements after this state is transferred from
R$^{}_{\textrm{a}}$ to R$^{}_{\textrm{b}}$ and then to
R$^{}_{\textrm{c}}$. Each transfer takes two steps, starting
with e.g. the state $| \textrm{Q}^{}_1
\textrm{R}^{}_{\textrm{a}} \textrm{R}^{}_{\textrm{b}} \rangle
{} = {} | \textrm{g} 2 0 \rangle$. The qubit is brought into
resonance with R$^{}_{\textrm{a}}$, and one photon is
Rabi-swapped to the qubit, at a rate $\sqrt{2}$ faster than the
usual one-photon rate~\cite{hofheinz:2008:a}, leaving the
system in the state $| \textrm{e} 1 0 \rangle$. The qubit is
then tuned into resonance with resonator R$^{}_{\textrm{b}}$
for a one-photon Rabi swap, resulting in the state $|
\textrm{g} 1 1 \rangle$
[Fig.~\ref{Figure3MatteoMariantoni201007}(iii)]. The qubit is
subsequently placed back in resonance with R$^{}_{\textrm{a}}$
for a one-photon swap, yielding $| \textrm{e} 0 1 \rangle$, and
brought into resonance with R$^{}_{\textrm{b}}$ to transfer the
second photon, ending with the state $| \textrm{g} 0 2 \rangle$
[Fig.~\ref{Figure3MatteoMariantoni201007}(iv)]. To finally
transfer the photons to resonator R$^{}_{\textrm{c}}$, the
process is repeated using qubit Q$^{}_2$, which completes the
full transfer of Fock state $| 2 \rangle$
[Fig.~\ref{Figure3MatteoMariantoni201007}(v)]. This process
resembles the well-known game `The towers of Hanoi', where a
set of disks with different diameters has to be moved between
three posts (the three resonators) while maintaining the larger
disks (Fock state $| 1 \rangle$, with the longer swapping time)
always at the bottom of each post, and the smaller disks (Fock
state $| 2 \rangle$, with shorter swapping time) on top.

\begin{figure}[t!]
    \centering
    \includegraphics[width=0.99\columnwidth]{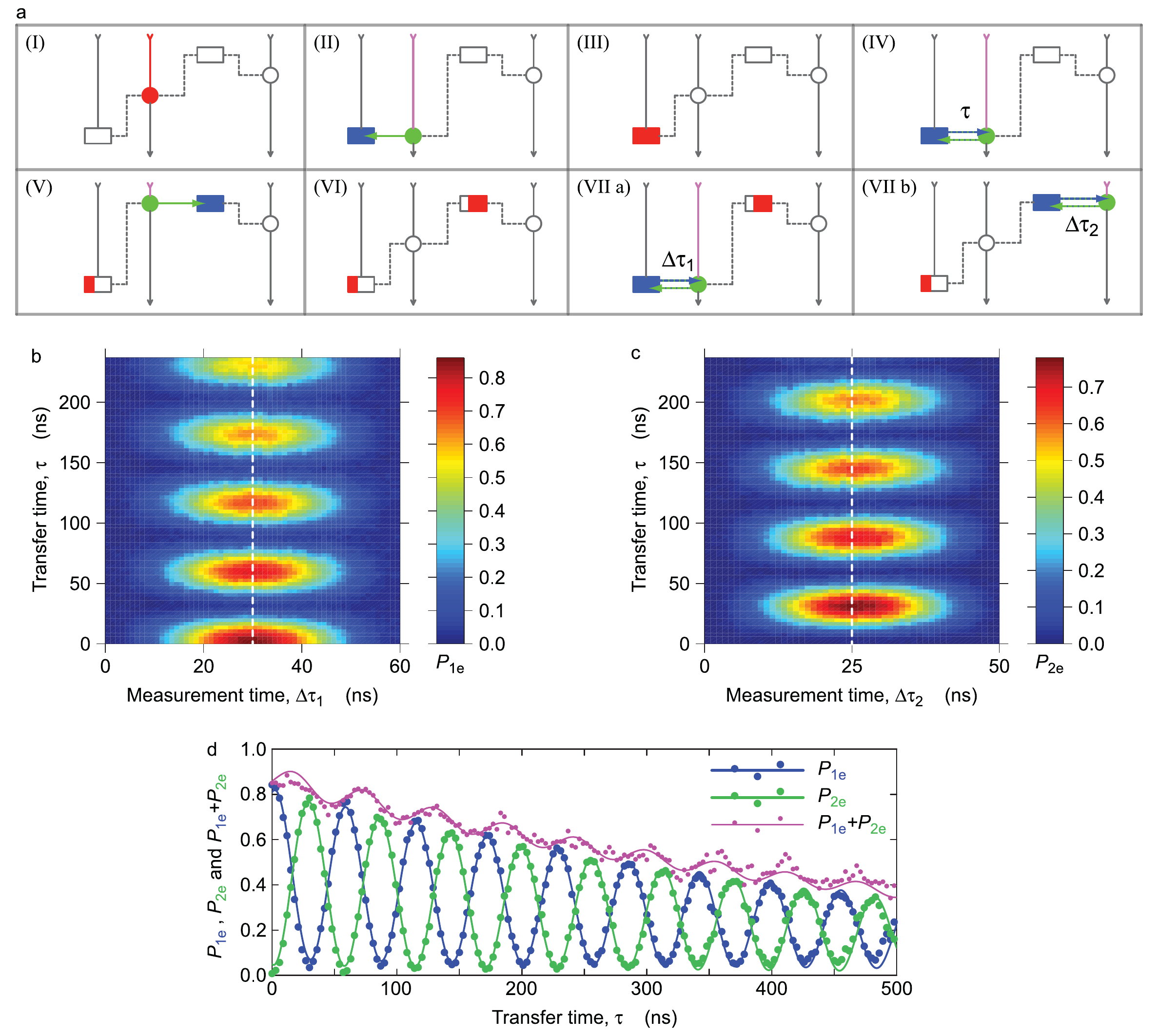}
    \caption{\footnotesize
\textbf{Two-resonator Rabi swaps.} \textbf{a,} Block diagram of
the preparation and measurement protocol. (I), Q$^{}_1$ is
excited by a $\pi$-pulse and, (II), brought into resonance with
R$^{}_{\textrm{a}}$ for a Rabi $\pi$-swap, at the end of which,
(III), R$^{}_{\textrm{a}}$ is populated by a one-photon Fock
state. In (IV), Q$^{}_1$ is brought into resonance with
R$^{}_{\textrm{a}}$ for a variable transfer time $\tau$ at the
end of which, (V), R$^{}_{\textrm{a}}$ is left partially
populated, while, (V) and (VI), the remaining energy is fully
transferred to R$^{}_{\textrm{b}}$ via a Rabi $\pi$-swap with
Q$^{}_1$. The probabilities for having the photon in
R$^{}_{\textrm{a}}$ or R$^{}_{\textrm{b}}$ can be varied
continuously by changing the transfer time $\tau$ (cf.~main
text). These probabilities are simultaneously measured for
R$^{}_{\textrm{a}}$ with Q$^{}_1$ (VII a) and for
R$^{}_{\textrm{b}}$ with Q$^{}_2$ (VII b). \textbf{b,} Qubit
probability $P^{}_{1 \textrm{e}}$ (colour scale bar) as a
function of the Q$^{}_1$-R$^{}_{\textrm{a}}$ measurement time
$\Delta \tau^{}_1$ (horizontal axis) and the transfer time
$\tau$ (vertical axis). \textbf{c,} Same as \textbf{b}, but for
Q$^{}_2$'s probability $P^{}_{2 \textrm{e}}$ as a function of
the Q$^{}_2$-R$^{}_{\textrm{b}}$ measurement time $\Delta
\tau^{}_2$. Measurement times are sufficient to display one
complete Rabi oscillation between the measurement qubit and
resonator, with a Rabi swap occurring at the center of each
horizontal axis (dashed white line), with multiple swaps
displayed as a function of the Q$^{}_1$-R$^{}_{\textrm{a}}$
transfer time $\tau$ (vertical direction). Cuts through the
probabilities along the dashed white lines (full Rabi swaps)
are shown in \textbf{d}, where the expected co-sinusoidal
oscillations are observed in Q$^{}_1$'s probability $P^{}_{1
\textrm{e}}$ (dark blue circles) (Q$^{}_2$'s probability
$P^{}_{2 \textrm{e}}$, light green circles), with the summed
probability $P {} = {} P^{}_{1 \textrm{e}} + P^{}_{2
\textrm{e}}$ (magenta circles) showing the expected slow decay
(cf.~main text). Solid lines are fits to data.
    }
    \label{Figure5MatteoMariantoni201007}
\end{figure}

A fundamental question for resonator-based quantum computing is
whether quantum states can be stored in a resonator and later
extracted and/or stored elsewhere. We demonstrate this
functionality in Fig.~\ref{Figure4MatteoMariantoni201007},
where a single photon is stored in each of the three resonators
for a variable time $\tau^{}_{\textrm{st}} / 3$ before being
transferred to the next resonator. Qubit measurements of the
resonator containing the photon display clear oscillations for
total storage times longer than $3\,\mu$s.
Figure~\ref{Figure4MatteoMariantoni201007}b is the same
experiment repeated with a two-photon Fock state. Clear
oscillations are visible for total storage times longer than
$1.5\,\mu$s. These experiments demonstrate the realization of a
programmable quantum information register.

We further show that we can generate quantum state entanglement
between the two resonators R$^{}_{\textrm{a}}$ and
R$^{}_{\textrm{b}}$. The protocol is diagrammed in
Fig.~\ref{Figure5MatteoMariantoni201007}a. A one-photon Fock
state is first stored in R$^{}_{\textrm{a}}$, placing the
system in the state $| \textrm{Q}^{}_1
\textrm{R}^{}_{\textrm{a}} \textrm{R}^{}_{\textrm{b}} \rangle
{} = {} | \textrm{g} 1 0 \rangle$
[Fig.~\ref{Figure5MatteoMariantoni201007}a(I)-(III)]. The qubit
is then used to perform a partial transfer of the photon to
R$^{}_{\textrm{b}}$, achieved by placing Q$^{}_1$ in resonance
with R$^{}_{\textrm{a}}$ and varying the transfer time $\tau$
[Fig.~\ref{Figure5MatteoMariantoni201007}a(IV)]. This leaves
the system in the entangled state $\alpha | \textrm{g} 1 0
\rangle + \beta | \textrm{e} 0 0 \rangle$, with
amplitudes~\cite{walls:2008:a} $\alpha {} = {} \cos ( \pi
g^{}_{1 \textrm{a}} \tau )$ and $\beta {} = {} i \sin( \pi
g^{}_{1 \textrm{a}} \tau )$. The qubit frequency $f^{}_1$ is
then tuned from $f^{}_{\textrm{a}}$ to $f^{}_{\textrm{b}}$, and
left there for a time equal to the Q$^{}_1$-R$^{}_{\textrm{b}}$
swap time, thus mapping the qubit state onto the resonator and
resulting in the two-resonator entangled state
    \setlength\arraycolsep{0pt}
\begin{displaymath}
\alpha | \textrm{g} 1 0 \rangle + \beta | \textrm{g} 0 1
\rangle {} = {} | \textrm{g} \rangle \otimes ( \alpha | 1 0
\rangle + \beta | 0 1 \rangle) \, .
\end{displaymath}

We then use both qubits to simultaneously measure the two
resonators, Q$^{}_1$ measuring R$^{}_{\textrm{a}}$ for a
measurement time $\Delta \tau^{}_1$ and Q$^{}_2$ measuring
R$^{}_{\textrm{b}}$ for a time $\Delta \tau^{}_2$.
Figure~\ref{Figure5MatteoMariantoni201007}b(c) shows the
resulting oscillations in $P^{}_{1 \textrm{e}}$ ($P^{}_{2
\textrm{e}}$) for Q$^{}_1$ (Q$^{}_2$) (colour bar scale), as a
function of the measurement time (horizontal axis) and
Q$^{}_1$-R$^{}_{\textrm{a}}$ transfer time $\tau$ (vertical
axis). The data display a single Rabi oscillation along the
horizontal axis, which would repeat if the measurement time
were increased, and also shows clear swaps as a function of the
transfer time $\tau$, as expected from the functional
dependence of $\alpha$ and $\beta$. If the measurement times
$\Delta \tau^{}_{1 , 2}$ are chosen to equal a full
qubit-resonator swap time (dashed white lines in
Fig.~\ref{Figure5MatteoMariantoni201007}b,c), the system will
be in the state $| \textrm{Q}^{}_1 \textrm{R}^{}_{\textrm{a}}
\textrm{R}^{}_{\textrm{b}} \textrm{Q}^{}_2 \rangle {} = {}
\alpha| \textrm{e} 0 0 \textrm{g} \rangle + \beta | \textrm{g}
0 0 \textrm{e} \rangle$. An ideal measurement of Q$^{}_1$ and
Q$^{}_2$ would then yield probabilities $P^{}_{1 \textrm{e}} {}
= {} | \alpha |^2_{} {} = {} \cos^2_{} ( \pi g^{}_{1
\textrm{a}} \tau )$ and $P^{}_{2 \textrm{e}} {} = {} | \beta
|^2_{} {} = {} \sin^2_{} ( \pi g^{}_{1 \textrm{a}} \tau )$ for
Q$^{}_1$ and Q$^{}_2$, respectively, as a function of the
transfer time $\tau$. In
Figure~\ref{Figure5MatteoMariantoni201007}d we display this
functional dependence, with a clear $180^{\circ}_{}$ phase
difference between the two probabilities and the summed
probability $P^{}_{1 \textrm{e}} + P^{}_{2 \textrm{e}}$ close
to unity, as expected. The probabilities decrease with $\tau$,
owing to the finite energy relaxation of Q$^{}_1$ and
R$^{}_{\textrm{a}}$. The decay time of R$^{}_{\textrm{b}}$ does
not contribute noticeably (cf.~Supplementary Information).
Fitting yields an effective two-resonator decay time
approximately equal to the harmonic mean of the relaxation
times $T^{\textrm{rel}}_1 {} \simeq {} 340$\,ns and
$T^{\textrm{rel}}_{\textrm{a}} {} \simeq {} 3.9\,\mu$s of
Q$^{}_1$ and R$^{}_{\textrm{a}}$, respectively,
$T^{\textrm{rel}}_{\textrm{ab}} {} \approx {} ( 1 / 2
T^{\textrm{rel}}_1 + 1 / 2 T^{\textrm{rel}}_{\textrm{a}} )^{-
1} {} \simeq {} 626$\,ns. We note that the phase qubit enables
a photon transfer between R$^{}_{\textrm{a}}$ and
R$^{}_{\textrm{b}}$ even though the two resonators are
separated in frequency by $\simeq {} 12000$ resonator
linewidths.

This last experiment demonstrates a true quantum version of the
`shell game', where the `pea' (the photon Fock state) is
simultaneously hidden under two shells, and the contestant's
selection of a shell constitutes a truly probabilistic
measurement. More generally, we have experimentally
demonstrated an architecture with three resonators and two
qubits that displays excellent quantum control over single and
double microwave photon Fock states. From a fundamental
perspective, these results demonstrate the potential of
multi-resonator circuit
QED~\cite{mariantoni:2008:a,helmer:2009:a}, both for scientific
study and for quantum information.

\vspace{5mm}


\section*{\textsf{\color{black}\large METHODS\\{\rule[6mm]{\columnwidth}{0.1mm}}}}

    {\footnotesize
 \textbf{Sample fabrication.}
The resonators are made from a $150$\,nm thick rhenium film
grown in a molecular-beam-epitaxy system onto a polished
sapphire substrate. All the other wiring layers are made from
sputtered aluminum with Al/AlO$^{}_x$/Al Josephson tunnel
junctions. All the micro-structures on the different layers are
patterned by means of optical lithography and etched by means
of inductively coupled plasma etching. Amorphous silicon is
used as dielectric insulator for the qubit shunting capacitors
and cross-overs. Our sample fabrication clearly shows the
flexibility offered by multi-layer processing.

The complete device is then bonded by aluminum wire-bonds to an
aluminum sample holder, which is bolted to the mixing chamber
of a dilution refrigerator operating at $\simeq {} 25$\,mK. A
detailed description of the fabrication techniques, electronics
and qubit calibration procedures can be found
elsewhere~\cite{hofheinz:2009:a}.
    }

 \vspace{2.5mm}

    {\footnotesize
 \textbf{Three-resonator circuit QED Hamiltonian.}
The Hamiltonian $\widehat{H}$ for the circuit of
Fig.~\ref{Figure1MatteoMariantoni201007}b can be written as the
sum of the Hamiltonians of each unit cell, $\widehat{H}^{}_1$
and $\widehat{H}^{}_2$. Neglecting the driving and dissipative
terms for simplicity, the Hamiltonian of the first circuit unit
cell can be expressed in the interaction picture with respect
to Q$^{}_1$ and R$^{}_{\textrm{a}}$ and R$^{}_{\textrm{b}}$ as
the combination of two Jaynes-Cummings interactions:
\begin{eqnarray}
\widehat{H}^{}_1 & {} = {} & h \frac{g^{}_{1 \textrm{a}}}{2} \left(
\hat{\sigma}^+_1 \hat{a}^{}_{} e^{+ i 2 \pi \Delta^{}_{1 {\rm
a}} t}_{} + \hat{\sigma}^-_1 \hat{a}^{\dag}_{} e^{- i 2 \pi
\Delta^{}_{1 {\rm a}} t}_{} \right) {} \nonumber\\[1.5mm]
& & {} + h \frac{g^{}_{1 \textrm{b}}}{2} \left(
\hat{\sigma}^+_1 \hat{b}^{}_{} e^{+ i 2 \pi \Delta^{}_{1 {\rm
b}} t}_{} + \hat{\sigma}^-_1 \hat{b}^{\dag}_{} e^{- i 2 \pi
\Delta^{}_{1 {\rm b}} t}_{} \right) \, ,
    \label{H:1}
\end{eqnarray}
where $\hat{\sigma}^{\pm}_1$ are the rising and lowering
operators for Q$^{}_1$, $\hat{a}^{}_{} , \hat{a}^{\dag}_{} ,
\hat{b}^{}_{}$ and $\hat{b}^{\dag}_{}$ the bosonic annihilation
and creation operators for R$^{}_{\textrm{a}}$ and
R$^{}_{\textrm{b}}$, respectively, and $\Delta^{}_{1
\textrm{a}} {} \equiv {} f^{}_1 - f^{}_{\textrm{a}}$ and
$\Delta^{}_{1 \textrm{b}} {} \equiv {} f^{}_1 -
f^{}_{\textrm{b}}$ the qubit-resonator de-tunings. The
Hamiltonian $\widehat{H}^{}_2$ for the second circuit unit cell
has an analogous expression.

In order to effectively switch off a particular qubit-resonator
Q$^{}_p$-R$^{}_q$ interaction, the condition $\Delta^{}_{p q}
{} \gg {} g^{}_{p q}$ must be fulfilled. This is the case when
Q$^{}_1$ (Q$^{}_2$) is at the idle point. On the contrary, when
$\Delta^{}_{p q} {} \rightarrow {} 0$, a resonant
Jaynes-Cummings interaction takes place enabling state
preparation and transfer in and between the resonators.
    }



\NPsection{Acknowledgements}

This work was supported by IARPA under ARO award
W911NF-08-1-0336 and under ARO award W911NF-09-1-0375. M.~M.
acknowledges support from an Elings Postdoctoral Fellowship.
Devices were made at the UC Santa Barbara Nanofabrication
Facility, a part of the NSF-funded National Nanotechnology
Infrastructure Network.\\


\NPsection{Author Contributions}

M.M. performed the experiments with the help of H.W. M.M.
analysed the data. M.M. and H.W. fabricated the sample. M.N.
provided software infrastructure. J.M.M. and E.L. designed the
custom electronics. All authors contributed to the fabrication
process, qubit design or experimental set-up, and discussed the
data analysis. M.M., J.M.M. and A.N.C. conceived the experiment
and co-wrote the paper.

\NPsection{Additional Information}

The authors declare no competing financial interests.
Supplementary information accompanies this paper.\\
Correspondence and requests for materials should be addressed
to A.N.C.\\

\newpage


\newcommand{\onlinecite}[1]{\hspace{-1 ex} \nocite{#1}\citenum{#1}}

\setcounter{figure}{0}

\noindent\parbox{\textwidth}{\flushleft\textsf{\color{black}\Huge
Supplementary information for `Photon shell game in
three-resonator circuit quantum electrodynamics'}}
    \vspace{3mm}

\noindent\parbox{\textwidth}{\flushleft
\textsf{\color{black}\Large
        Matteo~Mariantoni$^{\mathsf{1,3,*}}$,
        H.~Wang$^{1}$,
        Radoslaw~C.~Bialczak$^{1}$,
        M.~Lenander$^{1}$,
        Erik~Lucero$^{1}$,
        M.~Neeley$^1$,
        A.~D.~O'Connell$^{1}$,
        D.~Sank$^{1}$,
        M.~Weides$^{1}$,
        J.~Wenner$^{1}$,
        T.~Yamamoto$^{1,3}$,
        Y.~Yin$^{1}$,
        J.~Zhao$^1$,
        John~M.~Martinis$^{1,\dag}$
        and A.~N.~Cleland$^{1,\ddag}$
       }
    \vspace{2mm}

\textsf{\textbf{\small $^{1}$Department of Physics, University of California, Santa Barbara, California 93106, USA\\
$^{2}$Green Innovation Research Laboratories, NEC Corporation, Tsukuba, Ibaraki 305-8501, Japan\\
$^{3}$California NanoSystems Institute, University of California, Santa Barbara, California 93106, USA\\
$^{*}$e-mail: matmar@physics.ucsb.edu\\
$^{\dag}$e-mail: martinis@physics.ucsb.edu\\
$^{\ddag}$e-mail: anc@physics.ucsb.edu
    \vspace{5mm}
    }}}

\noindent\textsf{\small last updated: \today}\\
{\color{black}\rule[2mm]{\textwidth}{0.1mm}}

    {\noindent\textbf{
In this supplementary information, we provide further insight
into the transduction fidelity of microwave one- and two-photon
Fock states between three coplanar wave guide resonators. In
addition, we show that the exponential decay of the
two-resonator vacuum Rabi swaps fits very well with a simple
harmonic-mean decay model (confirmed by numerical simulations),
where both the qubit and resonator energy relaxation times
contribute to the effective decay of the two-resonator
dynamics.
    }

In order to realize three-resonator circuit quantum
electrodynamics (QED) experiments, a complex architecture must
be designed and fabricated on a single chip. In the main text
(cf.~Fig.~\ref{Figure1MatteoMariantoni201007}a) we have shown
an optical micrograph of the main circuit elements, comprising
three coplanar wave guide resonators and two superconducting
phase qubits. Figure~\ref{FigureS1MatteoMariantoni201007} shows
a detail of qubit Q$^{}_1$, together with its readout SQUID,
control and readout lines and the two coupling capacitors to
resonators R$^{}_{\textrm{a}}$ and R$^{}_{\textrm{b}}$.

\renewcommand{\figurename}{Figure Supplementary}
\renewcommand{\tablename}{Table Supplementary}

\begin{figure}[t!]
    \centering
    \includegraphics[width=0.99\columnwidth]{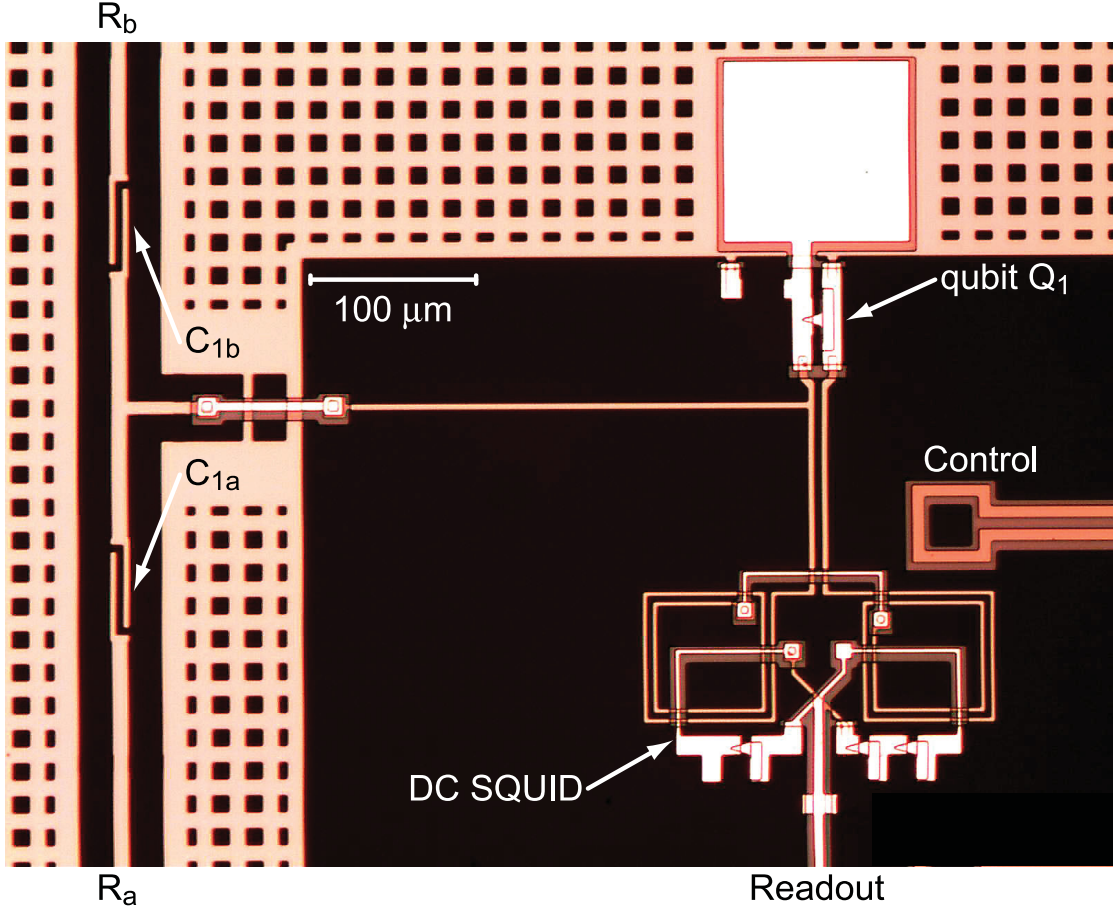}
    \caption{\footnotesize
\textbf{Detail of qubit Q$^{}_1$ coupled to resonators
R$^{}_{\textrm{a}}$ and R$^{}_{\textrm{b}}$.} The capacitors
$C^{}_{1 \textrm{a}}$ and $C^{}_{1 \textrm{b}}$ couple qubit
Q$^{}_1$ (gradiometer design) to resonators R$^{}_{\textrm{a}}$
and R$^{}_{\textrm{b}}$. The qubit state is read out by a DC
SQUID (gradiometer design). A portion of the control and
readout lines is also visible. A micrometer scale shows the
real circuit dimensions.
    }
    \label{FigureS1MatteoMariantoni201007}
\end{figure}

In order to unambiguously verify the high fidelity of transfer
of a one-photon Fock state from resonator R$^{}_{\textrm{a}}$
to resonator R$^{}_{\textrm{c}}$ via resonator
R$^{}_{\textrm{b}}$, we have also performed full-state Wigner
tomography on R$^{}_{\textrm{a}}$ and R$^{}_{\textrm{c}}$ for
the two prototypical examples of photon shell game of
Fig.~\ref{Figure2MatteoMariantoni201007}b(ii),(iv) (cf.~main
text). The results are displayed in
Fig.~S.~\ref{FigureS2MatteoMariantoni201007}, which shows the
measured Wigner functions $W ( \alpha )$ and corresponding
density matrices $\hat{\rho}$ for the Fock state $| 1 \rangle$
first stored in R$^{}_{\textrm{a}}$ and then in
R$^{}_{\textrm{c}}$ after transfer via R$^{}_{\textrm{b}}$.

The Wigner function is obtained following the steps explained
in Ref.~\onlinecite{hofheinz:2009:a:S}. In summary, the
resonator is first prepared in the desired microwave photon
state $| \Psi \rangle$. Second, the resonator is displaced by
injecting a coherent state $| - \alpha \rangle$ with complex
amplitude $\alpha {} = {} | \alpha | \exp( \varphi^{}_{\alpha}
)$, where $| \alpha |$ represents the coherent state real
amplitude and $\varphi^{}_{\alpha}$ its phase; the state is
injected through microwave control lines using a classical
pulsed source
(cf.~Fig.~\ref{Figure1MatteoMariantoni201007}a,b). Third, a
qubit in its energy ground state is brought into resonance with
the resonator for an interaction time long enough to execute
several qubit-resonator swaps. Fourth, a simple least-squares
fit of these oscillations allows us to obtain the probabilities
associated with the different photon number states, from which
it is finally possible to obtain the quasi-probability
distributions via the parity operator, giving access to
full-state Wigner
tomography~\cite{haroche:2006:a:S,leonhardt:1997:a:S,hofheinz:2009:a:S}.
The protocols for calibrating the amplitude and phase of the
coherent state used to displace the resonator are explained in
detail in Ref.~\onlinecite{hofheinz:2009:a:S}. From the Wigner
function it is possible to reconstruct the density matrix of
the photon state~\cite{leibfried:1996:a:S,bertet:2002:a:S} $|
\Psi \rangle$.

Given the theoretical and measured density matrices of a photon
state $| \Psi \rangle$, $\hat{\rho}^{}_{\textrm{th}}$ and
$\hat{\rho}^{}_{\textrm{m}}$, respectively, we define the state
fidelity as $\mathcal{F} {} \equiv {} \textrm{Tr} \{
\hat{\rho}^{}_{\textrm{th}} \, \hat{\rho}^{}_{\textrm{m}} \}$.
For the Fock state $| 1 \rangle$ prepared in
R$^{}_{\textrm{a}}$, with the associated Wigner function and
density matrix of
Fig.~S.~\ref{FigureS2MatteoMariantoni201007}a,c, we find
$\mathcal{F} {} \simeq {} 0.84$. This compares well with the
amplitude of the fit to the qubit-resonator swaps, which gives
a fidelity $\mathcal{F} {} \simeq {} 0.86$
(cf.~Fig.~\ref{Figure2MatteoMariantoni201007} in main text).
After being transferred to R$^{}_{\textrm{c}}$, the state has
the Wigner function and density matrix of
Fig.~\ref{FigureS2MatteoMariantoni201007}b,d and the
corresponding fidelity $\mathcal{F} {} \simeq {} 0.64$, which
is also consistent with the fidelity found with the simple
least-squares fit (cf.~Fig.~\ref{Figure2MatteoMariantoni201007}
in main text). The loss of fidelity occurring during the photon
transfer between the three resonators can be attributed to
qubit decoherence and slight calibration errors during the swap
qubit-resonator operations. Nevertheless, it is remarkable that
the density matrix associated with the state in
R$^{}_{\textrm{c}}$ is still very pure, with nearly negligible
spurious matrix elements and only a small contribution from the
$| 0 \rangle$ state.

\begin{figure}[t!]
    \centering
    \includegraphics[width=0.99\columnwidth]{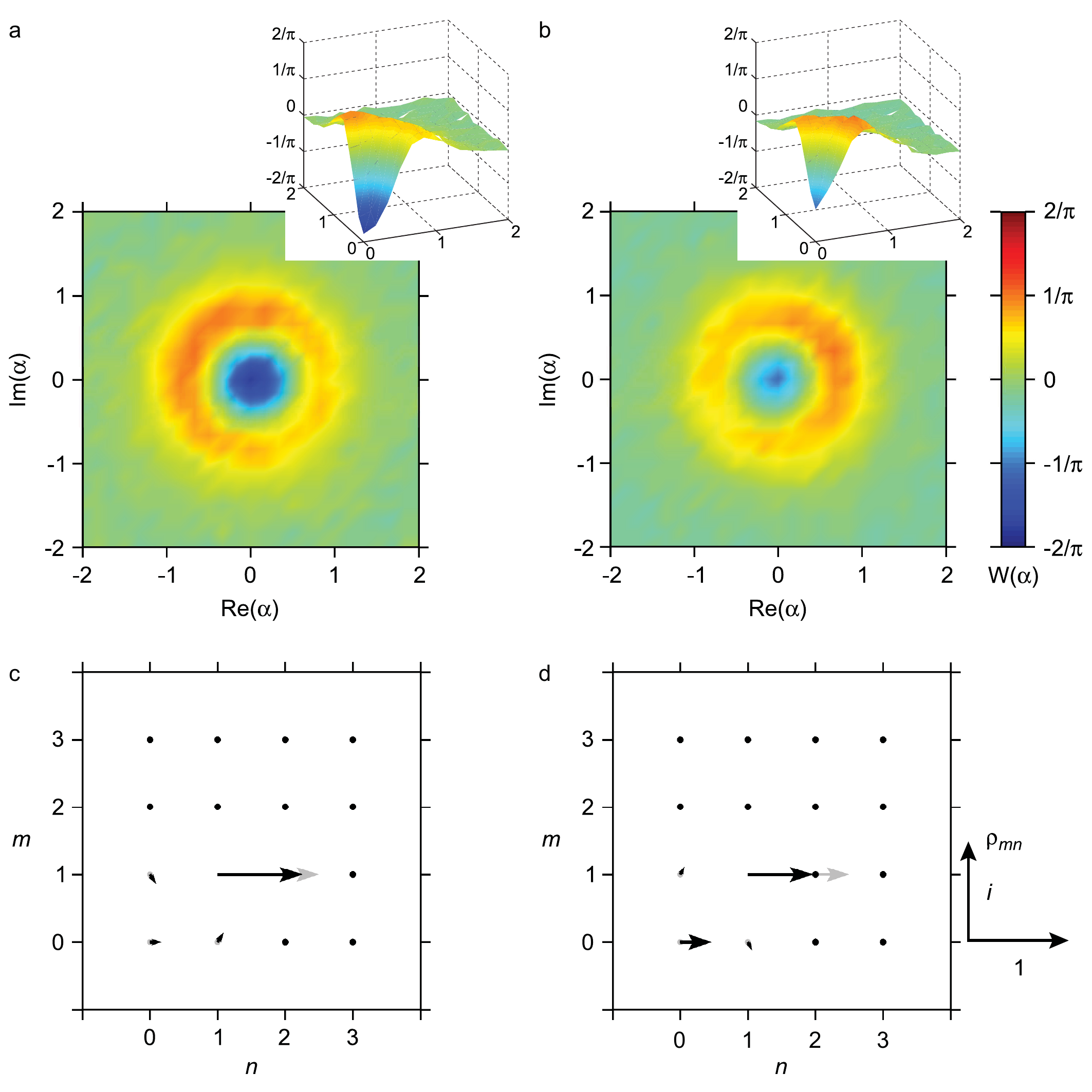}
    \caption{\footnotesize
\textbf{Wigner tomography for the photon shell game.}
\textbf{a,} Measured Wigner function $W ( \alpha )$ for
resonator R$^{}_{\textrm{a}}$ as a function of the complex
resonator amplitude $\alpha$ in square root of photon number
units (colour scale bar on the far right). The inset displays a
cut of the three-dimensional plot of the Wigner function.
\textbf{b,} Same as in \textbf{a}, but for resonator
R$^{}_{\textrm{c}}$ (colour scale bar on the right). Negative
quasi-probabilities clearly indicate the quantum-mechanical
nature of the intra-resonator states. \textbf{c,} Theoretical
(grey) and measured (black) values for the density matrix
associated with the state stored in resonator
R$^{}_{\textrm{a}}$, $\hat{\rho}$, projected onto the number
states $\rho^{}_{m n} {} \equiv {} \langle m | \hat{\rho} | n
\rangle$. The magnitude and phase of $\rho^{}_{m n}$ is
represented by the length and direction of an arrow in the
complex plane (the scale for the real and imaginary part is
reported on the far right). \textbf{d,} Same as in \textbf{c},
but for resonator R$^{}_{\textrm{c}}$. When representing the
density matrices, the resonator Hilbert space has been
truncated to the lowest four bosonic states.
    }
    \label{FigureS2MatteoMariantoni201007}
\end{figure}

Other two relevant figures of merit for the transfer of one-
and two-photon Fock states between resonators
R$^{}_{\textrm{a}}$, R$^{}_{\textrm{b}}$ and
R$^{}_{\textrm{c}}$ are represented by the harmonic purity of
the state (i.e. absence of beatings) and the $\sqrt{2}$ scaling
factor between swaps associated with the Fock
states~\cite{hofheinz:2008:a:S} $| 1 \rangle$ and $| 2
\rangle$. These properties can easily be estimated by computing
the fast Fourier transform (FFT) of the qubit-resonator vacuum
Rabi swaps.
Figure~Supplementary~\ref{FigureS3MatteoMariantoni201007} shows
the FFTs of the Rabi swaps for the key steps of the photon
shell game and `Towers of Hanoi', i.e. for a Fock $| 1 \rangle$
and $| 2 \rangle$ created in R$^{}_{\textrm{a}}$ and measured
with Q$^{}_1$, then transferred to R$^{}_{\textrm{b}}$ and
measured with Q$^{}_1$ and Q$^{}_2$, and finally transferred to
R$^{}_{\textrm{c}}$ and measured with Q$^{}_2$. The time-domain
swaps used to compute the FFTs are those in
Figs.~\ref{Figure2MatteoMariantoni201007}b and
\ref{Figure3MatteoMariantoni201007} in main text.

\begin{figure}[t!]
    \centering
    \includegraphics[width=0.99\columnwidth]{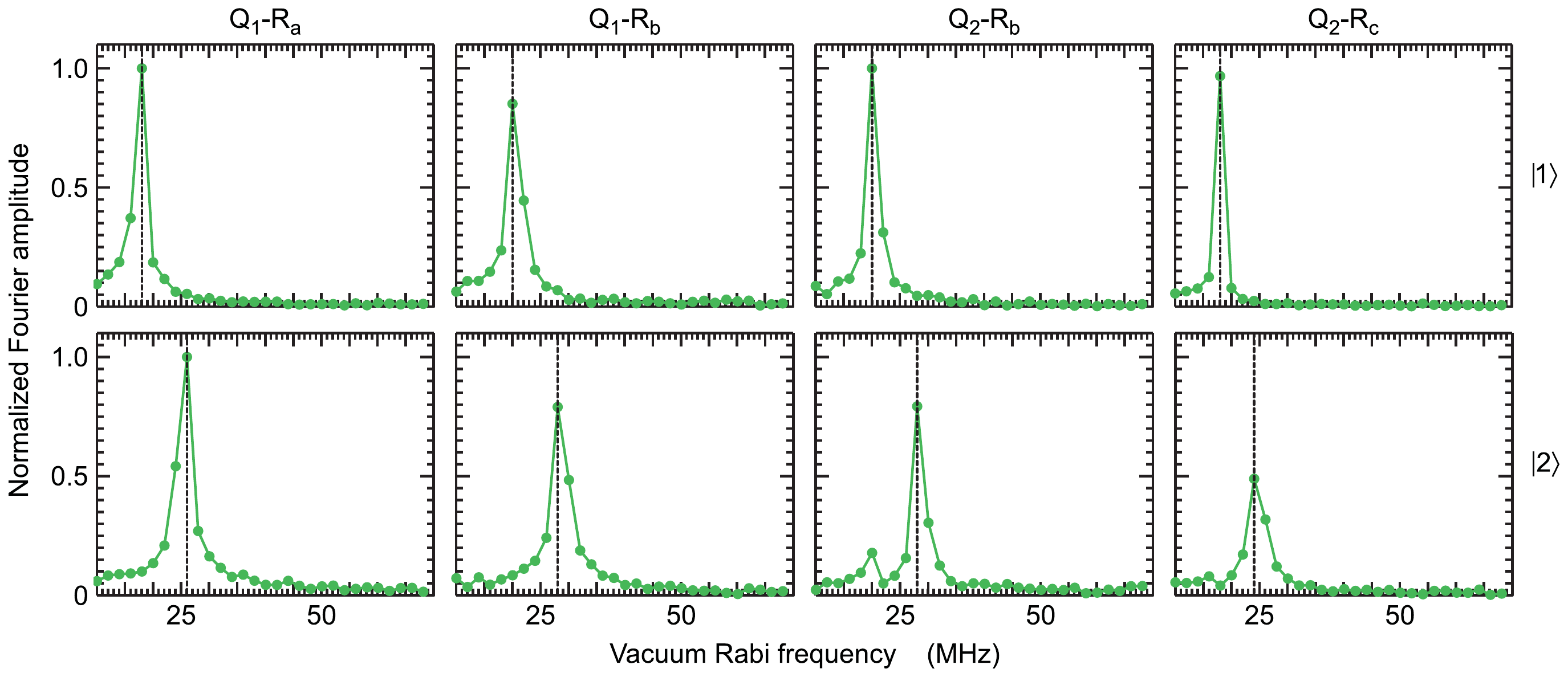}
    \caption{\footnotesize
\textbf{Fourier analysis for the photon shell game and `Towers
of Hanoi'.} Top sub-panels, normalized Fourier amplitude as a
function of vacuum Rabi frequency associated with the
qubit-resonator Rabi swaps for a one-photon Fock state $| 1
\rangle$. Bottom sub-panels, normalized Fourier amplitude for a
two-photon Fock state $| 2 \rangle$. Above each column is
indicated the respective qubit-resonator interaction. The
dashed black line in each sub-panel indicates the maximum
Fourier component. The amplitude re-normalization is calculated
with respect to the largest Fourier component for Fock state $|
1 \rangle$ (top sub-panels) and $| 2 \rangle$ (bottom
sub-panels), respectively. The $\sqrt{2}$
scaling~\cite{hofheinz:2009:a:S} between the $| 1 \rangle$ and
$| 2 \rangle$ states is clearly visible. The absence of
beatings (only a small beating, owing to some residual presence
of state $| 1 \rangle$, for state $| 2 \rangle$ in the
Q$^{}_2$-R$^{}_{\textrm{b}}$ sub-panel) shows the high level of
harmonic purity of the states transferred between the three
resonators, both for the photon shell game and for the more
complex `Towers of Hanoi'.
    }
    \label{FigureS3MatteoMariantoni201007}
\end{figure}

The last topic to be considered in this supplementary is the
energy relaxation mechanism of the two-resonator Rabi swaps of
Fig.~\ref{Figure5MatteoMariantoni201007}d in the main text.

In order to gain deeper insight into the decay dynamics of the
two-resonator Rabi, we first need information on the energy
relaxation times of qubit Q$^{}_1$ and resonators
R$^{}_{\textrm{a}}$ and R$^{}_{\textrm{b}}$.
Figure~Supplementary~\ref{FigureS4MatteoMariantoni201007} shows
the experimental measurement of these decay times. Note that
the resonators' energy relaxation time is determined by
preparing the resonator in a one-photon Fock state $| 1
\rangle$, and storing it for a variable time. After this time,
a qubit is brought on resonance with the resonator, the state
swapped into the qubit, and the qubit population then read
out~\cite{neeley:2008:a:S,wang:2008:a:S}.

The energy relaxation times obtained from a simple exponential
fit are reported in
Table~S.~\ref{TableS1MatteoMariantoni201007}. We use the qubit
and resonator energy relaxation times as well as the other
parameters listed in the table to solve numerically a
Lindblad-type master
equation~\cite{walls:2008:a:S,blais:2004:a:S}:
\begin{equation}
\dot{\hat{\rho}} {} = {} \frac{1}{i \hbar} ( \widehat{H}^{}_1
\hat{\rho} - \hat{\rho} \widehat{H}^{}_1 ) + \sum_{k {} = {} 1}^{3}
\, \hat{\mathcal{L}}^{}_k \, \hat{\rho} \, ,
    \label{two:resonator:master:equation}
\end{equation}
where $\dot{\hat{\rho}} {} \equiv {} ( \partial / \partial t )
\, \hat{\rho}$ is the time derivative of the total density
matrix $\hat{\rho}$ describing the system, $\widehat{H}^{}_1$
is the Hamiltonian of Eq.~(\ref{H:1}) (cf.~Methods' section in
main text), $\hat{\mathcal{L}}^{}_k$ is the Lindblad
superoperator defined as $\hat{\mathcal{L}}^{}_k \, \hat{\rho}
{} \equiv {} \gamma^{}_k ( \hat{X}^{}_k \hat{\rho}
\hat{X}^{\dag}_k - \hat{X}^{\dag}_k \hat{X}^{}_k \hat{\rho} / 2
- \hat{\rho} \hat{X}^{\dag}_k \hat{X}^{}_k / 2 )$ and $k {} \in
{} \mathbb{N}$. The qubit and resonators decay rates are
defined as $\gamma^{}_1 {} \equiv {} 1 /
T^{\textrm{rel}}_{\textrm{a}}$, $\gamma^{}_2 {} \equiv {} 1 /
T^{\textrm{rel}}_1$ and $\gamma^{}_3 {} \equiv {} 1 /
T^{\textrm{rel}}_{\textrm{b}}$ and the generating operators as
$\hat{X}^{}_1 {} \equiv {} \hat{a}^{}_{}$, $\hat{X}^{\dag}_1 {}
\equiv {} \hat{a}^{\dag}_{}$, $\hat{X}^{}_2 {} \equiv {}
\hat{\sigma}^{-}_{}$, $\hat{X}^{\dag}_2 {} \equiv {}
\hat{\sigma}^{+}_{}$, $\hat{X}^{}_3 {} \equiv {} \hat{b}^{}_{}$
and $\hat{X}^{\dag}_3 {} \equiv {} \hat{b}^{\dag}_{}$. We
numerically solve Eq.~(\ref{two:resonator:master:equation}) for
the pulse sequence shown in
Fig.~\ref{Figure5MatteoMariantoni201007}a (cf.~main text),
without accounting for the measurement process. The results are
displayed in Fig.~S.~\ref{FigureS5MatteoMariantoni201007},
where they are compared to the experimental data.

\begin{figure}[t!]
    \centering
    \includegraphics[width=0.36\columnwidth]{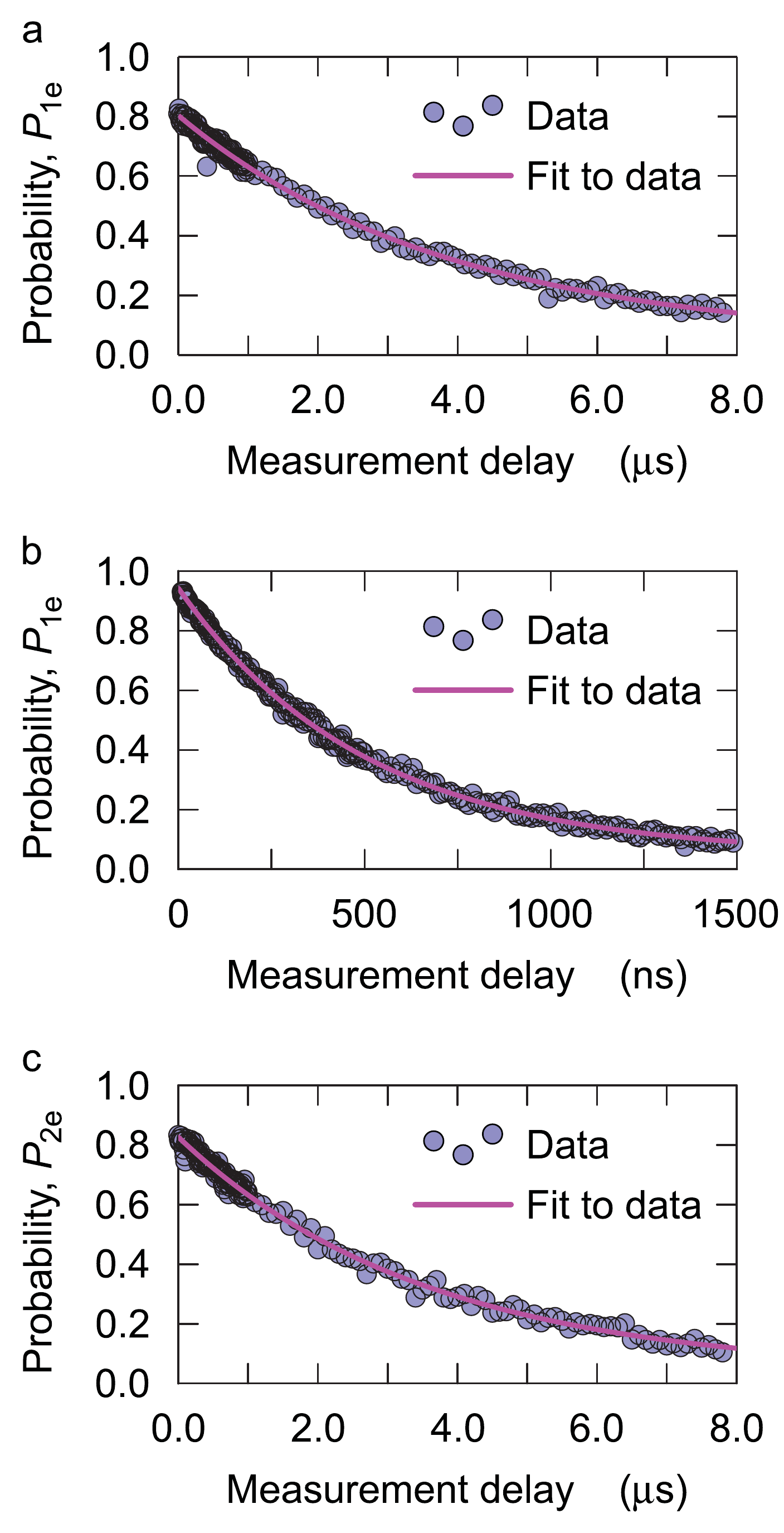}
    \caption{\footnotesize
\textbf{Energy relaxation for qubit Q$^{}_1$ and resonators
R$^{}_{\textrm{a}}$ and R$^{}_{\textrm{b}}$}. In all panels:
$P^{}_{p \textrm{e}}$ is the probability to find Q$^{}_p$ in $|
\textrm{e} \rangle$ as a function of measurement delay time
(i.e. the time a photon is stored in the resonator before being
swapped into the qubit). Full circles are data and solid
magenta lines exponential fits to data. \textbf{a,} Measurement
of the energy relaxation of resonator R$^{}_{\textrm{a}}$ using
qubit Q$^{}_1$ as a
detector~\cite{neeley:2008:a:S,wang:2008:a:S}. \textbf{b,}
Energy relaxation of qubit Q$^{}_1$. \textbf{c,} Same as in
\textbf{a}, but for resonator R$^{}_{\textrm{b}}$.
    }
    \label{FigureS4MatteoMariantoni201007}
\end{figure}

Figure~Supplementary~\ref{FigureS5MatteoMariantoni201007}a,c
shows the same data as in
Fig.~\ref{Figure5MatteoMariantoni201007}d, but for a transfer
time $\tau$ three times longer. Furthermore, the data shown in
Fig.~S.~\ref{FigureS5MatteoMariantoni201007}a,c were taken
using a different device compared to
Fig.~\ref{Figure5MatteoMariantoni201007}d, with longer qubit
relaxation times. The exponential decay obtained by the simple
harmonic mean model (cf.~main text) is superposed with the
data, making evident the qualitative validity of the model.
Figure~Supplementary~\ref{FigureS5MatteoMariantoni201007}b,d
shows the results of the simulations of
Eq.~(\ref{two:resonator:master:equation}) corresponding to the
experimental data of
Fig.~S.~\ref{FigureS5MatteoMariantoni201007}a,c, respectively.
Data and simulations are in very good agreement, supporting the
simple harmonic mean decay model. In particular, the
experimental decay time obtained by fitting the data is $\simeq
{} 840$\,ns, from simulations $\simeq {} 874$\,ns and from the
harmonic mean model $\simeq {} 896$\,ns. The amplitude of the
simulated two-resonator Rabi swaps was adjusted to the measured
amplitudes. The slight discrepancy between the experimental
data and simulations for the low region of the occupation
probabilities (causing an offset between data and simulations)
is because the simulations do not account for the measurement
process. Notice that we can safely assume that only qubit
Q$^{}_1$ and resonator R$^{}_{\textrm{a}}$, swapping for a
variable transfer time $\tau$, contribute to the effective
decay mechanism of the two-resonator Rabi dynamics. In fact,
the second resonator serves only as a mapping resonator, the
state of which is measured with Q$^{}_2$ typically after a time
$\Delta \tau^{}_2 {} \ll {} T^{\textrm{rel}}_{\textrm{b}}$
(cf.~main text and
Fig.~\ref{Figure5MatteoMariantoni201007}b,c). In other words,
examining the two-dimensional plots of
Fig.~\ref{Figure5MatteoMariantoni201007}b,c we expect two
distinct decay mechanisms. The first along the horizontal axis
($\Delta \tau^{}_1$ and $\Delta \tau^{}_2$). This decay is
practically negligible as this measurement is completed in
$\leqslant {} 30$\,ns. The second is along the vertical axis
($\tau$) related to the Q$^{}_1$-R$^{}_{\textrm{a}}$ swaps, as
explained above.

For a theoretical analysis of the decay mechanisms
characteristic of two-resonator dynamics in different regimes
we remind to Ref.~\onlinecite{reuther:2010:a:S}.

\begin{table*}[t!]
    \centering
\caption{\footnotesize \textbf{Main parameters used for the
numerical simulations of the two-resonator Rabi dynamics.}
$\delta$ represents the non-linearity of the superconducting
phase qubit, i.e. the energy difference in unit Hz between the
qubit transition frequencies relative to the ground state and
first excited state and to the first excited state and second
excited state, respectively~\cite{lucero:2010:a:S}. The
non-linearity has been used in the simulations to take into
account possible leakage outside of the qubit subspace.
$T^{\phi}_{\textrm{a}}$ and $T^{\phi}_{\textrm{b}}$ are the
dephasing times for resonators R$^{}_{\textrm{a}}$ and
R$^{}_{\textrm{b}}$, respectively. Since our main goal is to
understand the energy relaxation of the two-resonator Rabi
swaps, the qubit dephasing time $T^{\phi}_1$ has been neglected
in the simulations. All the other parameters are defined in the
main text.}
    \vspace{13.0pt}
 \begin{tabular}{c | c c c c c}
 \hline
 \hline
  \vspace{-1.5pt}
   \\
     \textbf{R$^{}_{\textrm{a}}$} \quad \quad & \quad \quad $f^{}_{\textrm{a}}$ \quad \quad & \quad \quad $-$
     \quad \quad & \quad \quad $g^{}_{1 \textrm{a}}$ \quad \quad & \quad \quad $T^{\textrm{rel}}_{\textrm{a}}$ \quad \quad & \quad \quad $T^{\phi}_{\textrm{a}}$
   \\
     {} \quad \quad & \quad \quad \footnotesize{(GHz)} \quad \quad & \quad \quad {}
     \quad \quad & \quad \quad \footnotesize{(MHz)} \quad \quad & \quad \quad \footnotesize{(ns)} \quad \quad & \quad \quad \footnotesize{(ns)}
   \\
     \vspace{-7.0pt}
   \\
     \hline
     \vspace{-7.0pt}
   \\
     {} \quad \quad & \quad \quad $6.340$ \quad \quad & \quad \quad {}
     \quad \quad & \quad \quad $17.95$ \quad \quad & \quad \quad $3881$ \quad \quad & \quad \quad $\gg {} T^{\textrm{rel}}_{\textrm{a}}$
   \\
     \vspace{-7.0pt}
   \\
     \hline
     \vspace{-7.0pt}
   \\
     \textbf{Q$^{}_1$} \quad \quad & \quad \quad $f^{}_1$ \quad \quad & \quad \quad $\delta$
     \quad \quad & \quad \quad $-$ \quad \quad & \quad \quad $T^{\textrm{rel}}_1$ \quad \quad & \quad \quad $T^{\phi}_{\textrm{a}}$
   \\
     {} \quad \quad & \quad \quad \footnotesize{(GHz)} \quad \quad & \quad \quad \footnotesize{(MHz)}
     \quad \quad & \quad \quad {} \quad \quad & \quad \quad \footnotesize{(ns)} \quad \quad & \quad \quad \footnotesize{(ns)}
   \\
     \vspace{-7.0pt}
   \\
     \hline
     \vspace{-7.0pt}
   \\
     {} \quad \quad & \quad \quad $6.563$ \quad \quad & \quad \quad $204.23$
     \quad \quad & \quad \quad {} \quad \quad & \quad \quad $507$ \quad \quad & \quad \quad $--$
   \\
     \vspace{-7.0pt}
   \\
     \hline
     \vspace{-7.0pt}
   \\
     \textbf{R$^{}_{\textrm{b}}$} \quad \quad & \quad \quad $f^{}_{\textrm{b}}$ \quad \quad & \quad \quad $-$
     \quad \quad & \quad \quad $g^{}_{1 \textrm{b}}$ \quad \quad & \quad \quad $T^{\textrm{rel}}_{\textrm{b}}$ \quad \quad & \quad \quad $T^{\phi}_{\textrm{b}}$
   \\
     {} \quad \quad & \quad \quad \footnotesize{(GHz)} \quad \quad & \quad \quad {}
     \quad \quad & \quad \quad \footnotesize{(MHz)} \quad \quad & \quad \quad \footnotesize{(ns)} \quad \quad & \quad \quad \footnotesize{(ns)}
   \\
     \vspace{-7.0pt}
   \\
     \hline
     \vspace{-7.0pt}
   \\
     {} \quad \quad & \quad \quad $6.815$ \quad \quad & \quad \quad {}
     \quad \quad & \quad \quad $20.25$ \quad \quad & \quad \quad $3549$ \quad \quad & \quad \quad $\gg {} T^{\textrm{rel}}_{\textrm{b}}$
   \\
     \vspace{-8.5pt}
   \\
   \hline
 \end{tabular}
    \label{TableS1MatteoMariantoni201007}
\end{table*}

\newpage

\begin{figure*}[p!]
    \centering
    \includegraphics[clip=,width=0.99\columnwidth]{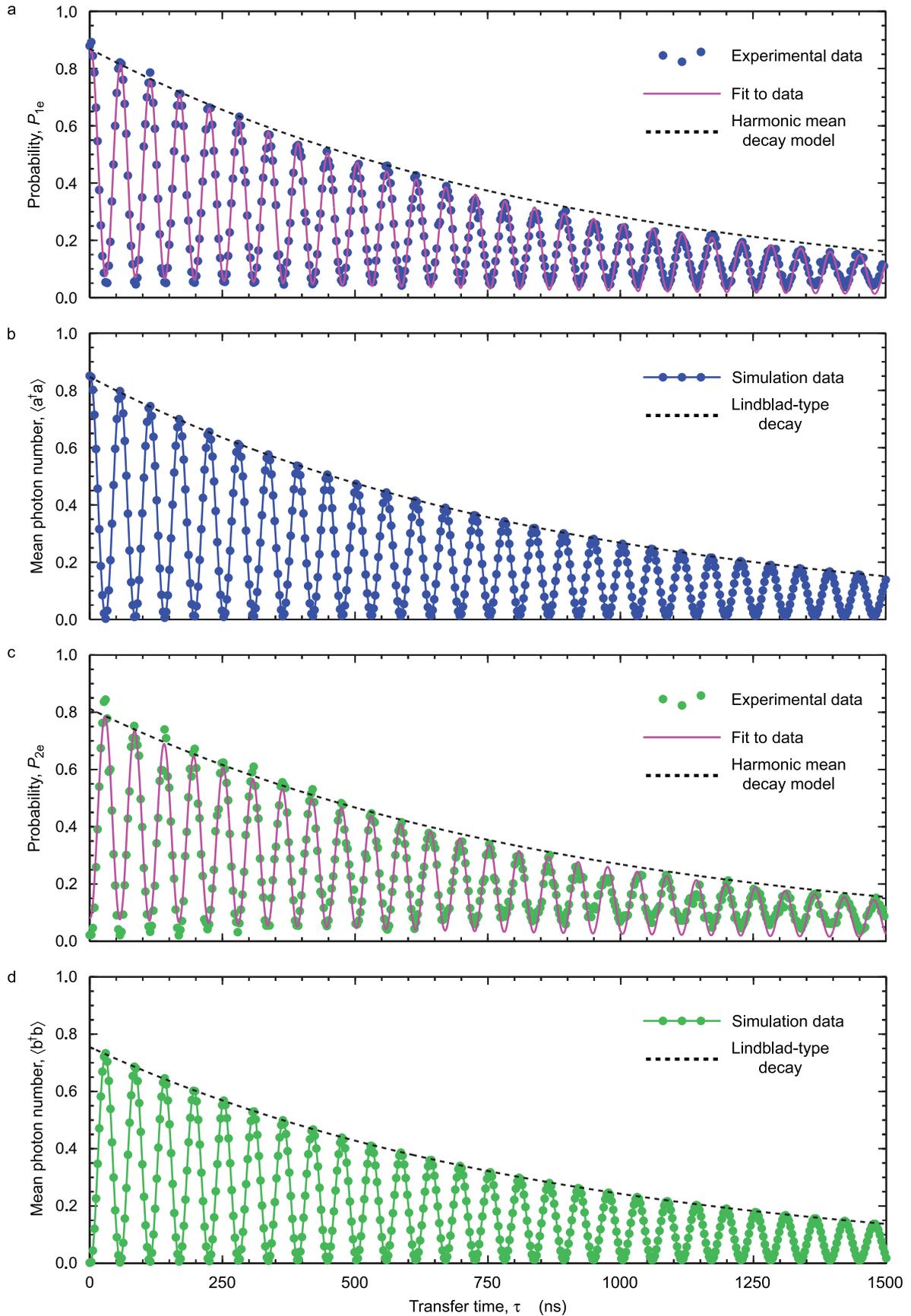}
    \caption{\footnotesize
\textbf{Comparison between experimental results and numerical
simulations of two-resonator Rabi swaps.} \textbf{a,}
R$^{}_{\textrm{a}}$ state dynamics measured as the probability
$P^{}_{1 \textrm{e}}$ to find Q$^{}_1$ in the excited state
versus transfer time $\tau$. Dark blue circles are data, solid
magenta line a least-squares fit to data and dashed black line
the exponential harmonic mean decay. \textbf{b,} Simulation
(parameters in Table~S.~\ref{TableS1MatteoMariantoni201007}) of
the data in \textbf{a}, showing the resonator mean photon
number $\langle \hat{a}^{\dag}_{} \hat{a}^{}_{} \rangle$ versus
$\tau$. The exponential decay envelope (dashed black line)
confirms the harmonic mean decay model. \textbf{c,} Same as in
\textbf{a}, but for R$^{}_{\textrm{b}}$ (cf.~main text). Light
green circles are data, solid magenta line a least-squares fit
to data and dashed black line the exponential harmonic mean
decay. \textbf{d,} Same as in \textbf{b}, but for
R$^{}_{\textrm{b}}$.
    }
    \label{FigureS5MatteoMariantoni201007}
\end{figure*}

As a last remark, it is worth mentioning that all data shown in
the main text and supplementary information were corrected for
measurement errors, following the standard procedure outlined
in Ref.~\onlinecite{ansmann:2009:a:S}.

In conclusion, in this supplementary information we have
validated the experimental results and conclusions presented in
the main text by means of full-state Wigner tomography, FFT
analysis and numerical simulations.

\newpage



\begin{thebibliography}{10}
\expandafter\ifx\csname url\endcsname\relax
  \def\url#1{\texttt{#1}}\fi
\expandafter\ifx\csname urlprefix\endcsname\relax\def\urlprefix{URL }\fi
\providecommand{\bibinfo}[2]{#2}
\providecommand{\eprint}[2][]{\url{#2}}

\bibitem{mabuchi:2002:a}
    \bibinfo{author}{Mabuchi, H.} \&
    \bibinfo{author}{Doherty, A.~C.}
\newblock
    \bibinfo{title}{Cavity Quantum Electrodynamics: Coherence in Context}.
\newblock
    \emph{\bibinfo{journal}{Science}}
    \textbf{\bibinfo{volume}{298}},
    \bibinfo{pages}{1372--1377}
    (\bibinfo{year}{2002}).

\bibitem{haroche:2006:a}
    \bibinfo{author}{Haroche, S.} \&
    \bibinfo{author}{Raimond, J.-M.}
\newblock
    \emph{\bibinfo{title}{Exploring the Quantum}}
    (\bibinfo{publisher}{Oxford University Press Inc.},
    \bibinfo{address}{New York},
    \bibinfo{year}{2006}).

\bibitem{walther:2006:a}
    \bibinfo{author}{Walther, H.},
    \bibinfo{author}{Varcoe, B.~T.~H.},
    \bibinfo{author}{Englert, B.-G.} \&
    \bibinfo{author}{Becker, T.}
\newblock
    \bibinfo{title}{Cavity Quantum Electrodynamics}.
\newblock
    \emph{\bibinfo{journal}{Rep.\ Prog.\ Phys.}}
    \textbf{\bibinfo{volume}{69}},
    \bibinfo{pages}{1325--1382}
    (\bibinfo{year}{2006}).

\bibitem{rauschenbeutel:2001:a}
    \bibinfo{author}{Rauschenbeutel, A.},
    \bibinfo{author}{Bertet, P.},
    \bibinfo{author}{Osnaghi, S.},
    \bibinfo{author}{Nogues, G.},
    \bibinfo{author}{Brune, M.},
    \bibinfo{author}{Raimond, J.-M.} \&
    \bibinfo{author}{Haroche, S.}
\newblock
    \bibinfo{title}{Controlled entanglement of two field modes
                    in a cavity quantum electrodynamics experiment}.
\newblock
    \emph{\bibinfo{journal}{Phys.\ Rev.\ A}}
    \textbf{\bibinfo{volume}{64}},
    \bibinfo{pages}{050301}
    (\bibinfo{year}{2001}).

\bibitem{gleyzes:2007:a}
    \bibinfo{author}{Gleyzes, S.},
    \bibinfo{author}{Kuhr, S.},
    \bibinfo{author}{Guerlin, C.},
    \bibinfo{author}{Bernu, J.},
    \bibinfo{author}{Del\'{e}glise, S.},
    \bibinfo{author}{Busk~Hoff, U.},
    \bibinfo{author}{Brune, M.},
    \bibinfo{author}{Raimond, J.-M.} \&
    \bibinfo{author}{Haroche, S.}
\newblock
    \bibinfo{title}{Quantum Jumps of Light Recording the Birth
                    and Death of a Photon in a Cavity}.
\newblock
    \emph{\bibinfo{journal}{Nature (London)}}
    \textbf{\bibinfo{volume}{446}},
    \bibinfo{pages}{297--300}
    (\bibinfo{year}{2007}).

\bibitem{deleglise:2008:a}
    \bibinfo{author}{Del\'{e}glise, S.},
    \bibinfo{author}{Dotsenko, I.},
    \bibinfo{author}{Sayrin, C.},
    \bibinfo{author}{Bernu, J.},
    \bibinfo{author}{Brune, M.},
    \bibinfo{author}{Raimond, J.-M.} \&
    \bibinfo{author}{Haroche, S.}
\newblock
    \bibinfo{title}{Reconstruction of non-classical cavity field states
                    with snapshots of their decoherence}.
\newblock
    \emph{\bibinfo{journal}{Nature (London)}}
    \textbf{\bibinfo{volume}{455}},
    \bibinfo{pages}{510--514}
    (\bibinfo{year}{2008}).

\bibitem{hijlkema:2007:a}
    \bibinfo{author}{Hijlkema, M.},
    \bibinfo{author}{Weber, B.},
    \bibinfo{author}{Specht, H.~P.},
    \bibinfo{author}{Webster, S.~C.},
    \bibinfo{author}{Kuhn, A.} \&
    \bibinfo{Rempe, G.}
\newblock
    \bibinfo{title}{A single-photon server with just one atom}.
    \newblock
    \emph{\bibinfo{journal}{Nature Physics}}
    \textbf{\bibinfo{volume}{3}},
    \bibinfo{pages}{253--255}
    (\bibinfo{year}{2007}).

\bibitem{wilk:2007:a}
    \bibinfo{author}{Wilk, T.},
    \bibinfo{author}{Webster, S.~C.},
    \bibinfo{author}{Kuhn, A.}, \&
    \bibinfo{author}{Rempe, G.}
\newblock
    \bibinfo{title}{Single-atom single-photon quantum interface}.
\newblock
    \emph{\bibinfo{journal}{Science}}
    \textbf{\bibinfo{volume}{317}},
    \bibinfo{pages}{488--490}
    (\bibinfo{year}{2007}).

\bibitem{dayan:2008:a}
    \bibinfo{author}{Dayan, B.},
    \bibinfo{author}{Parkins, A.~S.},
    \bibinfo{author}{Aoki, T.},
    \bibinfo{author}{Ostby, E.~P.},
    \bibinfo{author}{Vahala, K.~J.} \&
    \bibinfo{author}{Kimble, H.~J.}
\newblock
    \bibinfo{title}{A photon turnstile dynamically regulated by one atom}.
    \newblock
    \emph{\bibinfo{journal}{Science}}
    \textbf{\bibinfo{volume}{319}},
    \bibinfo{pages}{1062--1065}
    (\bibinfo{year}{2008}).

\bibitem{papp:2009:a}
    \bibinfo{author}{Papp, S.~B.},
    \bibinfo{author}{Soo~Choi, K.},
    \bibinfo{author}{Deng, H.},
    \bibinfo{author}{Lougovski, P.},
    \bibinfo{author}{van Enk, S.~J.} \&
    \bibinfo{author}{Kimble, H.~J.}
\newblock
    \bibinfo{title}{Characterization of multipartite entanglement
                    for one photon shared among four optical modes}.
\newblock
    \emph{\bibinfo{journal}{Science}}
    \textbf{\bibinfo{volume}{324}},
    \bibinfo{pages}{764--768}
    (\bibinfo{year}{2009}).

\bibitem{wallraff:2004:a}
    \bibinfo{author}{Wallraff, A.},
    \bibinfo{author}{Schuster, D.~I.},
    \bibinfo{author}{Blais, A.},
    \bibinfo{author}{Frunzio, L.},
    \bibinfo{author}{Huang, R.-S.},
    \bibinfo{author}{Majer, J.},
    \bibinfo{author}{Kumar, S.},
    \bibinfo{author}{Girvin, S.~M.} \&
    \bibinfo{author}{Schoelkopf, R.~J.}
\newblock
    \bibinfo{title}{Strong coupling of a single photon to a
                    superconducting qubit using circuit quantum electrodynamics}.
\newblock
    \emph{\bibinfo{journal}{Nature (London)}}
    \textbf{\bibinfo{volume}{431}},
    \bibinfo{pages}{162--167}
    (\bibinfo{year}{2004}).

\bibitem{chiorescu:2004:a}
    \bibinfo{author}{Chiorescu, I.},
    \bibinfo{author}{Bertet, P.},
    \bibinfo{author}{Semba, K.},
    \bibinfo{author}{Nakamura, Y.},
    \bibinfo{author}{Harmans, C.~J.~P.~M.} \&
    \bibinfo{author}{Mooij, J.~E.}
\newblock
    \bibinfo{title}{Coherent dynamics of a flux qubit
                    coupled to a harmonic oscillator}.
\newblock
    \emph{\bibinfo{journal}{Nature (London)}}
    \textbf{\bibinfo{volume}{431}},
    \bibinfo{pages}{159--162}
    (\bibinfo{year}{2004}).

\bibitem{johansson:2006:a}
    \bibinfo{author}{Johansson, J.},
    \bibinfo{author}{Saito, S.},
    \bibinfo{author}{Meno, T.},
    \bibinfo{author}{Nakano, H.},
    \bibinfo{author}{Ueda, M.},
    \bibinfo{author}{Semba, K.} \&
    \bibinfo{author}{Takayanagi, H.}
\newblock
    \bibinfo{title}{Vacuum Rabi oscillations in a macroscopic
                    superconducting qubit {LC} oscillator system}.
\newblock
    \emph{\bibinfo{journal}{Phys.\ Rev.\ Lett.}}
    \textbf{\bibinfo{volume}{96}},
    \bibinfo{pages}{127006}
    (\bibinfo{year}{2006}).

\bibitem{schoelkopf:2008:a}
    \bibinfo{author}{Schoelkopf, R.~J.} \&
    \bibinfo{author}{Girvin, S.~M.}
\newblock
    \bibinfo{title}{Wiring up quantum systems}.
\newblock
    \emph{\bibinfo{journal}{Nature (London)}}
    \textbf{\bibinfo{volume}{451}},
    \bibinfo{pages}{664--669}
    (\bibinfo{year}{2008}).

\bibitem{makhlin:2001:a}
    \bibinfo{author}{Makhlin, {\relax Yu}.},
    \bibinfo{author}{Sch\"{o}n, G.} \&
    \bibinfo{author}{Shnirman, A.}
\newblock
    \bibinfo{title}{Quantum-state engineering with Josephson-junction devices}.
\newblock
    \emph{\bibinfo{journal}{Rev.\ Mod.\ Phys.}}
    \textbf{\bibinfo{volume}{73}},
    \bibinfo{pages}{357--400}
    (\bibinfo{year}{2001}).

\bibitem{wendin:2006:a}
    \bibinfo{author}{Wendin, G.} \&
    \bibinfo{author}{Shumeiko, V.~S.}
\newblock
    in \emph{\bibinfo{booktitle}{Handbook of Theoretical and
             Computational Nanotechnology}}
    Vol.~\bibinfo{volume}{3}
    (eds.~\bibinfo{editor}{Rieth, M.} \&
    \bibinfo{editor}{Schommers, W.})
    \bibinfo{pages}{223--309}
    (\bibinfo{publisher}{American Scientific Publishers},
    \bibinfo{address}{Los Angeles},
    \bibinfo{year}{2006}).

\bibitem{you:2005:a}
    \bibinfo{author}{You, J.~Q.} \&
    \bibinfo{author}{Nori, F.}
\newblock
    \bibinfo{title}{Quantum-state engineering with Josephson-junction devices}.
\newblock
    \emph{\bibinfo{journal}{Phys.\ Today}}
    \textbf{\bibinfo{volume}{58 (11)}},
    \bibinfo{pages}{42--47}
    (\bibinfo{year}{2005}).

\bibitem{clarke:2008:a}
    \bibinfo{author}{Clarke, J.} \&
    \bibinfo{author}{Wilhelm, F.~K.}
\newblock
    \bibinfo{title}{Superconducting quantum bits}.
\newblock
    \emph{\bibinfo{journal}{Nature (London)}}
    \textbf{\bibinfo{volume}{453}},
    \bibinfo{pages}{1031--1042}
    (\bibinfo{year}{2008}).

\bibitem{houck:2007:a}
    \bibinfo{author}{Houck, A.~A.},
    \bibinfo{author}{Schuster, D.~I.},
    \bibinfo{author}{Gambetta, J.~M.},
    \bibinfo{author}{Schreier, J.~A.},
    \bibinfo{author}{Johnson, B.~R.},
    \bibinfo{author}{Chow, J.~M.}
    \bibinfo{author}{Frunzio, L.},
    \bibinfo{author}{Majer, J.}
    \bibinfo{author}{Devoret, M.~H.},
    \bibinfo{author}{Girvin, S.~M.} \&
    \bibinfo{author}{Schoelkopf, R.~J.}
\newblock
    \bibinfo{title}{Generating single microwave photons in a circuit}.
\newblock
    \emph{\bibinfo{journal}{Nature (London)}}
    \textbf{\bibinfo{volume}{449}},
    \bibinfo{pages}{328--331}
    (\bibinfo{year}{2007}).

\bibitem{sillanpaa:2007:a}
    \bibinfo{author}{Sillanp\"{a}\"{a}, M.~A.},
    \bibinfo{author}{Park, J.~I.} \&
    \bibinfo{author}{Simmonds, R.~W.}
\newblock
    \bibinfo{title}{Coherent quantum state storage and transfer
                    between two phase qubits via a resonant cavity}.
\newblock
    \emph{\bibinfo{journal}{Nature (London)}}
    \textbf{\bibinfo{volume}{449}},
    \bibinfo{pages}{438--442}
    (\bibinfo{year}{2007}).

\bibitem{hofheinz:2008:a}
    \bibinfo{author}{Hofheinz, M.},
    \bibinfo{author}{Weig, E.~M.},
    \bibinfo{author}{Ansmann, M.},
    \bibinfo{author}{Bialczak, R.~C.}
    \bibinfo{author}{Lucero, E.},
    \bibinfo{author}{Neeley, M.}
    \bibinfo{author}{O'Connell, A.~D.},
    \bibinfo{author}{Wang, H.}
    \bibinfo{author}{Martinis, J.~M.} \&
    \bibinfo{author}{Cleland, A.~N.}
\newblock
    \bibinfo{title}{Generation of Fock states in a superconducting quantum circuit}.
\newblock
    \emph{\bibinfo{journal}{Nature (London)}}
    \textbf{\bibinfo{volume}{454}},
    \bibinfo{pages}{310--314}
    (\bibinfo{year}{2008}).

\bibitem{deppe:2008:a}
    \bibinfo{author}{Deppe, F.},
    \bibinfo{author}{Mariantoni, M.},
    \bibinfo{author}{Menzel, E.~P.},
    \bibinfo{author}{Marx, A.}
    \bibinfo{author}{Saito, S.},\
    \bibinfo{author}{Kakuyanagi, K.}
    \bibinfo{author}{Tanaka, H.},
    \bibinfo{author}{Meno, T.}
    \bibinfo{author}{Semba, K.}
    \bibinfo{author}{Takayanagi, H.}
    \bibinfo{author}{Solano, E.} \&
    \bibinfo{author}{Gross, R.}
\newblock \bibinfo{title}{Two-photon probe of the Jaynes–Cummings model and
                          controlled symmetry breaking in circuit QED}.
\newblock \emph{\bibinfo{journal}{Nature Phsys.}}
    \textbf{\bibinfo{volume}{4}},
    \bibinfo{pages}{686--691}
    (\bibinfo{year}{2008}).

\bibitem{hofheinz:2009:a}
    \bibinfo{author}{Hofheinz, M.},
    \bibinfo{author}{Wang, H.},
    \bibinfo{author}{Ansmann, M.},
    \bibinfo{author}{Bialczak, R.~C.}
    \bibinfo{author}{Lucero, E.},
    \bibinfo{author}{Neeley, M.}
    \bibinfo{author}{O'Connell, A.~D.},
    \bibinfo{author}{Sank, D.}
    \bibinfo{author}{Sank, D.},
    \bibinfo{author}{Wenner, J.}
    \bibinfo{author}{Martinis, J.~M.} \&
    \bibinfo{author}{Cleland, A.~N.}
\newblock
    \bibinfo{title}{Synthesizing arbitrary quantum states
                    in a superconducting resonator}.
\newblock
    \emph{\bibinfo{journal}{Nature (London)}}
    \textbf{\bibinfo{volume}{459}},
    \bibinfo{pages}{546--549}
    (\bibinfo{year}{2009}).

\bibitem{wang:2009:a}
    \bibinfo{author}{Wang, H.},
    \bibinfo{author}{Hofheinz, M.},
    \bibinfo{author}{Ansmann, M.},
    \bibinfo{author}{Bialczak, R.~C.}
    \bibinfo{author}{Lucero, E.},
    \bibinfo{author}{Neeley, M.},
    \bibinfo{author}{O'Connell, A.~D.},
    \bibinfo{author}{Sank, D.},
    \bibinfo{author}{Weides, M.},
    \bibinfo{author}{Wenner, J.},
    \bibinfo{author}{Cleland, A.~N.} \&
    \bibinfo{author}{Martinis, J.~M.}
\newblock
    \bibinfo{title}{Decoherence dynamics of complex photon states
                    in a superconducting circuit}.
\newblock
    \emph{\bibinfo{journal}{Phys.\ Rev.\ Lett.}}
    \textbf{\bibinfo{volume}{103}},
    \bibinfo{pages}{3200404}
    (\bibinfo{year}{2009}).

\bibitem{leek:2010:a}
    \bibinfo{author}{Leek, P.~J.},
    \bibinfo{author}{Baur, M.},
    \bibinfo{author}{Fink, J.~M.},
    \bibinfo{author}{Bianchetti, R.}
    \bibinfo{author}{Steffen, L.},
    \bibinfo{author}{Filipp, S.} \&
    \bibinfo{author}{Wallraff, A.}
\newblock
    \bibinfo{title}{Cavity QED with separate photon storage
                    and qubit readout modes}.
\newblock
    \emph{\bibinfo{journal}{Phys.\ Rev.\ Lett.}}
    \textbf{\bibinfo{volume}{104}},
    \bibinfo{pages}{100504}
    (\bibinfo{year}{2010}).

\bibitem{johnson:2010:a}
    \bibinfo{author}{Johnson, B.~R.},
    \bibinfo{author}{Reed, M.~D.},
    \bibinfo{author}{Houck, A.~A.},
    \bibinfo{author}{Schuster, D.~I.}
    \bibinfo{author}{Bishop, L.~S.},
    \bibinfo{author}{Ginossar, E.}
    \bibinfo{author}{Gambetta, J.~M.},
    \bibinfo{author}{DiCarlo, L.}
    \bibinfo{author}{Frunzio, L.},
    \bibinfo{author}{Girvin, S.~M.} \&
    \bibinfo{author}{Schoelkopf, R.~J.}
\newblock
    \bibinfo{title}{Quantum non-demolition detection
                    of single microwave photons in a circuit}.
\newblock
    \emph{\bibinfo{journal}{Nature Physics}}
    \textbf{\bibinfo{volume}{DOI: 10.1038/NPHYS1710}},
    (\bibinfo{year}{2010}).

\bibitem{mariantoni:2008:a}
    \bibinfo{author}{Mariantoni, M.},
    \bibinfo{author}{Deppe, F.},
    \bibinfo{author}{Marx, A.},
    \bibinfo{author}{Gross, R.}
    \bibinfo{author}{Wilhelm, F.~K.},
    \& \bibinfo{author}{Solano, E.}
\newblock
    \bibinfo{title}{Two-resonator circuit quantum electrodynamics:
                    A superconducting quantum switch}.
\newblock
    \emph{\bibinfo{journal}{Phys.\ Rev.\ B}}
    \textbf{\bibinfo{volume}{78}},
    \bibinfo{pages}{104508}
    (\bibinfo{year}{2008}).

\bibitem{sun:2005:a}
    \bibinfo{author}{Sun, C.~P.},
    \bibinfo{author}{Wei, L.~F.},
    \bibinfo{author}{Liu, Y.~X.} \&
    \bibinfo{author}{Nori, F.}
\newblock
    \bibinfo{title}{Quantum transducers: integrating transmission lines and
                    nanomechanical resonators via charge qubits}.
\newblock
    \emph{\bibinfo{journal}{Phys.\ Rev.\ A}}
    \textbf{\bibinfo{volume}{73}},
    \bibinfo{pages}{022318}
    (\bibinfo{year}{2006}).

\bibitem{helmer:2009:a}
    \bibinfo{author}{Helmer, F.},
    \bibinfo{author}{Mariantoni, M.},
    \bibinfo{author}{Fowler, A.~G.},
    \bibinfo{author}{von Delft, J.}
    \bibinfo{author}{Solano, E.} \&
    \bibinfo{author}{Marquardt, F.}
\newblock
    \bibinfo{title}{Cavity grid for scalable quantum computation
                    with superconducting circuits}.
\newblock
    \emph{\bibinfo{journal}{Eurphys.\ Lett.}}
    \textbf{\bibinfo{volume}{85}},
    \bibinfo{pages}{50007}
    (\bibinfo{year}{2009}).

\bibitem{walls:2008:a}
    \bibinfo{author}{Walls, D.~F.} \&
    \bibinfo{author}{Milburn, G.~J.}
\newblock
    \bibinfo{title}{\textit{Quantum Optics} 2nd ed.}
    (\bibinfo{publisher}{Springer},
    \bibinfo{address}{Berlin-Heidelberg},
    \bibinfo{year}{2008}).

\end{thebibliography}

\begin{thebibliography}{10}
\expandafter\ifx\csname url\endcsname\relax
  \def\url#1{\texttt{#1}}\fi
\expandafter\ifx\csname
urlprefix\endcsname\relax\def\urlprefix{URL }\fi
\providecommand{\bibinfo}[2]{#2}
\providecommand{\eprint}[2][]{\url{#2}}

\bibitem{hofheinz:2009:a:S}
    \bibinfo{author}{Hofheinz, M.},
    \bibinfo{author}{Wang, H.},
    \bibinfo{author}{Ansmann, M.},
    \bibinfo{author}{Bialczak, R.~C.}
    \bibinfo{author}{Lucero, E.},
    \bibinfo{author}{Neeley, M.}
    \bibinfo{author}{O'Connell, A.~D.},
    \bibinfo{author}{Sank, D.}
    \bibinfo{author}{Sank, D.},
    \bibinfo{author}{Wenner, J.}
    \bibinfo{author}{Martinis, J.~M.} \&
    \bibinfo{author}{Cleland, A.~N.}
\newblock
    \bibinfo{title}{Synthesizing arbitrary quantum states
                    in a superconducting resonator}.
\newblock
    \emph{\bibinfo{journal}{Nature (London)}}
    \textbf{\bibinfo{volume}{459}},
    \bibinfo{pages}{546--549}
    (\bibinfo{year}{2009}).

\bibitem{haroche:2006:a:S}
    \bibinfo{author}{Haroche, S.} \&
    \bibinfo{author}{Raimond, J.-M.}
\newblock
    \emph{\bibinfo{title}{Exploring the Quantum}}
    (\bibinfo{publisher}{Oxford University Press Inc.},
    \bibinfo{address}{New York},
    \bibinfo{year}{2006}).

\bibitem{leonhardt:1997:a:S}
    \bibinfo{author}{Leonhardt, U.}
\newblock
    \emph{\bibinfo{title}{Measuring the Quantum State of Light}}
    (\bibinfo{publisher}{Cambridge University Press},
    \bibinfo{address}{Cambridge},
    \bibinfo{year}{1997}).

\bibitem{leibfried:1996:a:S}
    \bibinfo{author}{Leibfried, D.},
    \bibinfo{author}{Meekhof, D.~M.},
    \bibinfo{author}{King, B.~E.},
    \bibinfo{author}{Monroe, C.},
    \bibinfo{author}{Itano, W.~M.} \&
    \bibinfo{author}{Wineland, D.~J.}
\newblock
    \bibinfo{title}{Experimental determination of the motional
                    quantum state of a trapped atom}.
\newblock
    \emph{\bibinfo{journal}{Phys.\ Rev.\ Lett.}}
    \textbf{\bibinfo{volume}{77}},
    \bibinfo{pages}{4281–-4285}
    (\bibinfo{year}{1996}).

\bibitem{bertet:2002:a:S}
    \bibinfo{author}{Bertet, P.},
    \bibinfo{author}{Auffeves, A.},
    \bibinfo{author}{Maioli, P.},
    \bibinfo{author}{Osnaghi, S.},
    \bibinfo{author}{Meunier, T.},
    \bibinfo{author}{Brune, M.},
    \bibinfo{author}{Raimond, J.-M.} \&
    \bibinfo{author}{Haroche, S.}
\newblock
    \bibinfo{title}{Direct measurement of the Wigner function
                    of a one-photon Fock state in a cavity}.
\newblock
    \emph{\bibinfo{journal}{Phys.\ Rev.\ Lett.}}
    \textbf{\bibinfo{volume}{89}},
    \bibinfo{pages}{200402}
    (\bibinfo{year}{2002}).

\bibitem{hofheinz:2008:a:S}
    \bibinfo{author}{Hofheinz, M.},
    \bibinfo{author}{Weig, E.~M.},
    \bibinfo{author}{Ansmann, M.},
    \bibinfo{author}{Bialczak, R.~C.}
    \bibinfo{author}{Lucero, E.},
    \bibinfo{author}{Neeley, M.}
    \bibinfo{author}{O'Connell, A.~D.},
    \bibinfo{author}{Wang, H.}
    \bibinfo{author}{Martinis, J.~M.} \&
    \bibinfo{author}{Cleland, A.~N.}
\newblock
    \bibinfo{title}{Generation of Fock states in a superconducting quantum circuit}.
\newblock
    \emph{\bibinfo{journal}{Nature (London)}}
    \textbf{\bibinfo{volume}{454}},
    \bibinfo{pages}{310--314}
    (\bibinfo{year}{2008}).

\bibitem{neeley:2008:a:S}
    \bibinfo{author}{Neeley, M.},
    \bibinfo{author}{Ansmann, M.},
    \bibinfo{author}{Bialczak, R.~C.}
    \bibinfo{author}{Hofheinz, M.}
    \bibinfo{author}{Katz, N.}
    \bibinfo{author}{Lucero, E.},
    \bibinfo{author}{O'Connell, A.},
    \bibinfo{author}{Wang, H.},
    \bibinfo{author}{Cleland, A.~N.} \&
    \bibinfo{author}{Martinis, J.~M.}
\newblock
    \bibinfo{title}{Process tomography of quantum memory in a
                    Josephson-phase qubit coupled to a two-level state}.
\newblock
    \emph{\bibinfo{journal}{Nature Physics}}
    \textbf{\bibinfo{volume}{4}},
    \bibinfo{pages}{523--526}
    (\bibinfo{year}{2008}).

\bibitem{wang:2008:a:S}
    \bibinfo{author}{Wang, H.},
    \bibinfo{author}{Hofheinz, M.},
    \bibinfo{author}{Ansmann, M.},
    \bibinfo{author}{Bialczak, R.~C.}
    \bibinfo{author}{Lucero, E.},
    \bibinfo{author}{Neeley, M.}
    \bibinfo{author}{O'Connell, A.~D.},
    \bibinfo{author}{Sank, D.}
    \bibinfo{author}{Wenner, J.},
    \bibinfo{author}{Cleland, A.~N.} \&
    \bibinfo{author}{Martinis, J.~M.}
\newblock
    \bibinfo{title}{Measurement of the decay of Fock states
                    in a superconducting quantum circuit}.
\newblock
    \emph{\bibinfo{journal}{Phys.\ Rev.\ Lett.}}
    \textbf{\bibinfo{volume}{101}},
    \bibinfo{pages}{240401}
    (\bibinfo{year}{2008}).

\bibitem{walls:2008:a:S}
    \bibinfo{author}{Walls, D.~F.} \&
    \bibinfo{author}{Milburn, G.~J.}
\newblock
    \bibinfo{title}{\textit{Quantum Optics} 2nd ed.}
    (\bibinfo{publisher}{Springer},
    \bibinfo{address}{Berlin-Heidelberg},
    \bibinfo{year}{2008}).

\bibitem{blais:2004:a:S}
    \bibinfo{author}{Blais, A.},
    \bibinfo{author}{Huang, R.-S.},
    \bibinfo{author}{Wallraff, A.},
    \bibinfo{author}{Girvin, S.~M.} \&
    \bibinfo{author}{Schoelkopf, R.~J.}
\newblock
    \bibinfo{title}{Cavity quantum electrodynamics for
                    superconducting electrical circuits:
                    An architecture for quantum computation}.
\newblock
    \emph{\bibinfo{journal}{Phys.\ Rev.\ A}}
    \textbf{\bibinfo{volume}{69}},
    \bibinfo{pages}{062320}
    (\bibinfo{year}{2004}).

\bibitem{ansmann:2009:a:S}
    \bibinfo{author}{Ansmann, M.},
    \bibinfo{author}{Wang, H.},
    \bibinfo{author}{Bialczak, R.~C.},
    \bibinfo{author}{Hofheinz, M.},
    \bibinfo{author}{Lucero, E.},
    \bibinfo{author}{Neeley, M.},
    \bibinfo{author}{O'Connell, A.~D.},
    \bibinfo{author}{Sank, D.},
    \bibinfo{author}{Weides, M.},
    \bibinfo{author}{Wenner, J.},
    \bibinfo{author}{Cleland, A.~N.} \&
    \bibinfo{author}{Martinis, J.~M.}
\newblock
    \bibinfo{title}{Violation of Bell's inequality
                    in Josephson phase qubits}.
\newblock
    \emph{\bibinfo{journal}{Nature (London)}}
    \textbf{\bibinfo{volume}{461}},
    \bibinfo{pages}{504--506}
    (\bibinfo{year}{2009}).

\bibitem{lucero:2010:a:S}
    \bibinfo{author}{Lucero, E.},
    \bibinfo{author}{Kelly, J.},
    \bibinfo{author}{Bialczak, R.~C.},
    \bibinfo{author}{Lenander, M.},
    \bibinfo{author}{Mariantoni, M.},
    \bibinfo{author}{Neeley, M.},
    \bibinfo{author}{O'Connell, A.~D.},
    \bibinfo{author}{Sank, D.},
    \bibinfo{author}{Wang, H.},
    \bibinfo{author}{Weides, M.},
    \bibinfo{author}{Wenner, J.},
    \bibinfo{author}{Yamamoto, T.},
    \bibinfo{author}{Cleland, A.~N.} \&
    \bibinfo{author}{Martinis, J.~M.}
\newblock
    \bibinfo{title}{Reduced phase error through optimized
                    control of a superconducting qubit}.
\newblock
    \emph{\bibinfo{journal}{eprint arXiv}}
    \textbf{\bibinfo{volume}{1007.1690}},
    (\bibinfo{year}{2010}).

\bibitem{reuther:2010:a:S}
    \bibinfo{author}{Reuther, G.~M.},
    \bibinfo{author}{Zueco, D.},
    \bibinfo{author}{Deppe, F.},
    \bibinfo{author}{Hoffmann, E.},
    \bibinfo{author}{Menzel, E.~P.},
    \bibinfo{author}{Wei{\ss}l, T.},
    \bibinfo{author}{Mariantoni, M.},
    \bibinfo{author}{Kohler, S.},
    \bibinfo{author}{Marx, A.},
    \bibinfo{author}{Solano, E.},
    \bibinfo{author}{Gross, R.}, \&
    \bibinfo{author}{H\"{a}nggi, P.}
\newblock
    \bibinfo{title}{Two-resonator circuit quantum electrodynamics:
                    dissipative theory}.
\newblock
    \emph{\bibinfo{journal}{Phys.\ Rev.\ B}}
    \textbf{\bibinfo{volume}{81}},
    \bibinfo{pages}{144510}
    (\bibinfo{year}{2010}).

\end{thebibliography}
\end{document}